\documentclass[journal]{IEEEtran}

\usepackage{xcolor}
\usepackage{amsfonts}
\usepackage{amssymb}
\usepackage{bm}
\usepackage{bbm}
\usepackage{mathrsfs}
\usepackage{relsize}
\usepackage{mathtools}
\usepackage{algorithm}
\usepackage{algpseudocode}
\usepackage[colorlinks,urlcolor=blue,linkcolor=blue,citecolor=blue,breaklinks=true]{hyperref}
\usepackage{url}
\usepackage{subfigure}
\usepackage{amsmath}
\DeclareMathOperator*{\argmax}{argmax}
\usepackage[nocompress]{cite}
\ifCLASSINFOpdf
  \usepackage[pdftex]{graphicx}
\else
\fi

\begin{document}

\title{Two-Timescale Optimization Framework for IAB-Enabled Heterogeneous UAV Networks}

%
%
% author names and IEEE memberships
% note positions of commas and nonbreaking spaces ( ~ ) LaTeX will not break
% a structure at a ~ so this keeps an author's name from being broken across
% two lines.
% use \thanks{} to gain access to the first footnote area
% a separate \thanks must be used for each paragraph as LaTeX2e's \thanks
% was not built to handle multiple paragraphs
%

\author{Jikang~Deng,~\IEEEmembership{Student Member,~IEEE,} Hui~Zhou,~\IEEEmembership{Member,~IEEE,} and Mohamed-Slim~Alouini,~\IEEEmembership{Fellow,~IEEE}
		
		\thanks{Jikang~Deng, and Mohamed-Slim~Alouini are with CEMSE Division, King Abdullah University of Science and Technology (KAUST), Thuwal, 23955-6900, Kingdom of Saudi
Arabia (KSA) (email: \href{mailto:jikang.deng@kaust.edu.sa}{jikang.deng@kaust.edu.sa};
\href{mailto:slim.alouini@kaust.edu.sa}{slim.alouini@kaust.edu.sa})}
		\thanks{Hui Zhou is with Centre for Future Transport and Cities, Coventry University, U.K. (email:\href{mailto:hui.zhou@coventry.ac.uk}{hui.zhou@coventry.ac.uk}). This work was done while he was working at King Abdullah University of Science and Technology.}
		
	}% <-this % stops a space

% make the title area
\maketitle

% As a general rule, do not put math, special symbols or citations
% in the abstract or keywords.
\begin{abstract}
In post-disaster scenarios, the rapid deployment of adequate communication infrastructure is essential to support disaster search, rescue, and recovery operations. To achieve this, uncrewed aerial vehicle (UAV) has emerged as a promising solution for emergency communication due to its low cost and deployment flexibility. However, conventional untethered UAV (U-UAV) is constrained by size, weight, and power (SWaP) limitations, making it incapable of maintaining the operation of a macro base station. To address this limitation, we propose a heterogeneous UAV-based framework that integrates tethered UAV (T-UAV) and U-UAVs, where U-UAVs are utilized to enhance the throughput of cell-edge ground user equipments (G-UEs) and guarantee seamless connectivity during G-UEs' mobility to safe zones. It is noted that the integrated access and backhaul (IAB) technique is adopted to support the wireless backhaul of U-UAVs. Accordingly, we formulate a two-timescale joint user scheduling and trajectory control optimization problem, aiming to maximize the downlink throughput under asymmetric traffic demands and G-UEs' mobility. To solve the formulated problem, we proposed a two-timescale multi-agent deep deterministic policy gradient (TTS-MADDPG) algorithm based on the centralized training and distributed execution paradigm. Numerical results show that the proposed algorithm outperforms other benchmarks, including the two-timescale multi-agent proximal policy optimization (TTS-MAPPO) algorithm and MADDPG scheduling method, with robust and higher throughput. Specifically, the proposed algorithm obtains up to 12.2\% average throughput gain compared to the MADDPG scheduling method.
\end{abstract}

% Note that keywords are not normally used for peerreview papers.
\IEEEpeerreviewmaketitle

\section{Introduction}
\IEEEPARstart{N}{ext-generation} wireless communications networks are expected to provide higher capacity, enhanced reliability, and ubiquitous connectivity \cite{nguyen2021integrated}. In post-disaster scenarios, such as the aftermath of flooding, hurricanes, or earthquakes, the demand for persistent and reliable communication networks to support search, rescue, and recovery becomes critical. However, deploying efficient fixed terrestrial base station (TBS) systems in these scenarios poses considerable challenges due to terrain damage and widespread power outages. More importantly, a key component of emergency response is the rapid establishment of safe zones to ensure the well-being of affected populations. As disaster victims naturally move toward these safe zones, maintaining seamless and adaptive communication services becomes essential \cite{deepak2019overview}. In such cases, fixed TBSs, due to the lack of flexibility, are inadequate in providing reliable connections for mobile ground user equipments (G-UEs) in post-disaster scenarios. These limitations underscore the need for more adaptable and resilient communication solutions to support efficient emergency response and disaster relief activities.

Recently, the non-terrestrial network (NTN) has been identified as an important research direction for solving the above challenges, where diverse NTN platforms, including uncrewed aerial vehicle (UAV) \cite{matracia2021topological}, high-altitude platforms (HAPs) \cite{liu2023joint,shang2024enhancing}, and satellites \cite{xiao2024space,deng2025orthogonality}, can be deployed in various scenarios based on their unique characteristics. Among these platforms, UAV-based wireless networks stand out as a promising solution for post-disaster emergency communication, owing to their inherent advantages of high mobility, low cost, and flexible deployment \cite{licea2024robotics,zhou2021real}. Despite the significant advantages of UAV-based networks, UAVs as aerial base stations still face several challenges, such as the limited battery capacity and loading capability, which limit their wide adoption in practice. For example, the loading capability of a typical DJI untethered-UAV (U-UAV) is 2.7 kg with 31 minutes of flight time, while the typical weight of a macro base station (BS) is over 15 kg with power consumption around 3.8 kWh \cite{deng2025distributed}.
To solve the practical deployment issue above, tethered UAV (T-UAV) has been regarded as a promising solution to facilitate the deployment of a macro UAV-based network by leveraging its enhanced loading capability and tethered system \cite{zhang2021tethered}.

However, T-UAV is typically tethered via fiber-optic cables and power lines to ensure stable backhaul and sufficient power supply, which significantly restricts its mobility. The existing works on T-UAV for emergency communication overlook the limited mobility of T-UAV, which leads to degraded communication performance at the cell-edge UEs.
More importantly, the G-UEs in the disaster area are required to move toward specific safety zones, where the T-UAV cannot guarantee seamless connectivity to mobile G-UEs due to its limited mobility. Therefore, by leveraging the advantages of T-UAV and U-UAV, we propose a novel heterogeneous UAV network consisting of both T-UAV and U-UAVs, where T-UAV serves as a macro BS with stable backhaul and U-UAVs serve as micro BSs with high mobility.

\begin{figure}[ht]
  \centering
  \includegraphics[width = 8.2cm]{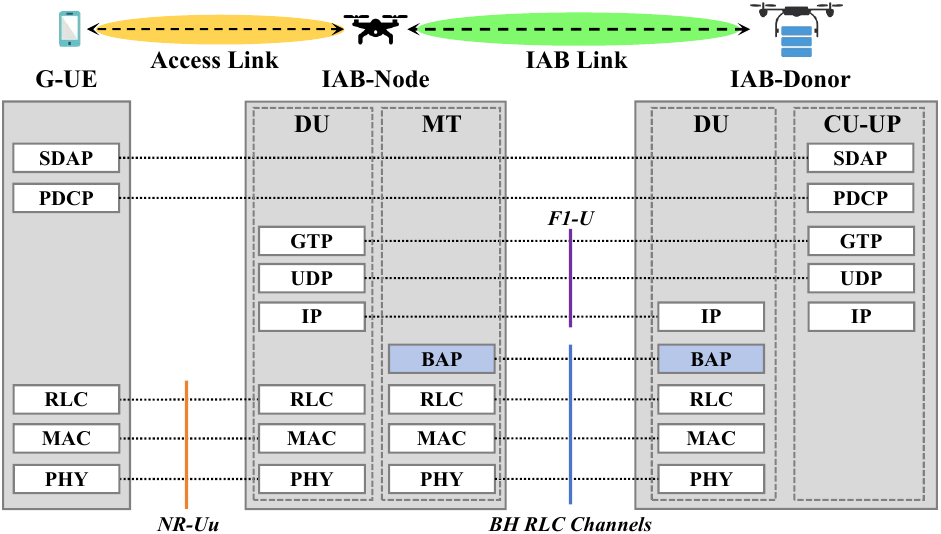}
  \caption{IAB framework in UAV network.}
  \label{IAB_Sturcture}
\end{figure}

To overcome the U-UAVs' backhaul limitations, the integrated access and backhaul (IAB) technique, promoted by the 3rd Generation Partnership Project (3GPP), has emerged as a promising solution \cite{madapatha2020integrated}. As shown in Fig. \ref{IAB_Sturcture}. The IAB-Donor
is defined as the BS providing the connections between G-UEs and the core network while also providing wireless backhauling capabilities to IAB-Nodes. The IAB-Node refers to a BS that enables wireless access for G-UEs while also wirelessly backhauling the associated access traffic. Specifically, each IAB-Node is equipped with a distributed unit (DU) and a mobile termination (MT), where the MT establishes connections with the IAB-Donor, and the DU establishes connections to G-UEs. The IAB-Donor is also equipped with a DU to provide connections for G-UEs and MTs of downstream IAB-Nodes.

Several studies have explored the application of IAB in UAV networks \cite{fouda2019interference,sheng2024enabling,diamanti2021prospect,zhang2024energy} to enhance backhaul connectivity. In \cite{fouda2019interference}, the authors leveraged UAVs as hovering IAB-Nodes and TBS as the IAB-Donor to provide backhaul links, and aimed at improving the interference management in this IAB network. 
In \cite{sheng2024enabling}, the authors proposed a mobility adaptable IAB scheme for coverage enhancement in an IAB network with fixed TBSs to provide dynamic backhaul for UAV-based aerial base stations.
These approaches for backhauling have inherent limitations, as conventional TBSs are typically down-tilted towards the ground to serve G-UEs, which cannot provide reliable IAB connections to UAVs. In \cite{diamanti2021prospect}, the authors proposed a game theory-based mechanism to optimize the energy efficiency of uplink transmission in a reconfigurable intelligent surface (RIS) assisted IAB network with a fixed-position UAV, which neglects the optimization of UAV trajectory to make full use of the UAV's mobility. In \cite{zhang2024energy}, the authors designed a combination of deep reinforcement
learning (DRL) and convex optimization techniques to jointly optimize the UAV's trajectory and resource allocation policy. However, these works either focus on the aerial-to-ground (A2G) IAB link or assume static G-UEs for simplicity, which cannot capture the characteristics of emergency communication in the post-disaster scenario effectively. More importantly, the optimal scheduling policy adapting to asymmetric traffic demands over the access links and the IAB links has not been investigated yet.

Recently, some studies have applied DRL and multi-agent deep reinforcement learning (MADRL) algorithms to enhance multi-UAV-assisted communications \cite{liu2018energy,zhong2021multi,ding20203d,guo2022multi,zhou2024joint}. 
The authors in \cite{liu2018energy} employed DRL based on the centralized training and execution (CTE) framework to optimize UAVs' trajectories, which leads to high information sharing overheads for large-scale networks with high-dimensional observations.
Alternatively, the authors in \cite{zhong2021multi,ding20203d} employed the decentralized training and execution (DTE) framework, where the authors in \cite{zhong2021multi} focused on UAVs' trajectory and power allocation optimization based on deep Q-network (DQN) algorithm, and the authors in \cite{ding20203d} optimized UAVs' trajectory design and band allocation with the deep deterministic policy gradient (DDPG) algorithm. However, the DTE framework fails to deal with the non-stationarity challenge due to the lack of coordination and inefficient exploration under partial observability.

To address the scalability and coordination challenges in multi-UAV networks, the centralized training and distributed execution (CTDE) framework is proposed and adopted by some existing works \cite{guo2022multi,zhou2024joint}. Specifically, in \cite{guo2022multi}, the authors optimized joint trajectory and power control in non-orthogonal multiple access (NOMA) enabled UAV communications by multi-agent deep deterministic policy gradient (MADDPG) to minimize transmission latency. In \cite{zhou2024joint}, the authors proposed a heterogeneous coordinated QMIX (HC-QMIX) algorithm to optimize UAV trajectories, user association, and transmit power in a multi-UAV emergency communication system. However, the above learning-based solutions mainly focused on optimizing the UAV trajectory and power allocation, without considering the G-UEs' mobility and asymmetric traffic demands under the IAB setting.

Motivated by the limitations of existing works above, this work focuses on designing an algorithm to jointly optimize user scheduling and trajectory control in a heterogeneous UAV-based emergency communication network, aiming to maximize the downlink successfully transmitted throughput under G-UEs' mobility. The main contributions of this paper are as follows:
\begin{itemize}
    \item We propose a novel IAB-enabled heterogeneous UAV-based emergency communication network for post-disaster scenarios. Specifically, T-UAV (i.e., IAB-Donor) provides connections to both the associated G-UEs and U-UAVs (i.e., IAB-Nodes) based on a stable backhaul connection to the core network. The IAB-Nodes dynamically serve the associated cell-edge G-UEs in the disaster area, and provide seamless communication service while G-UEs move towards the safe zones.

    \item We first formulate a downlink throughput maximization problem to jointly optimize the user scheduling and trajectory control of UAVs, subject to scheduling and velocity constraints. We then propose a two-timescale MADDPG (TTS-MADDPG) algorithm based on the CTDE framework to solve the formulated mixed-integer nonlinear programming (MINLP) problem. Specifically, each U-UAV aims to optimize the user scheduling and trajectory control policy using local actor networks, where the scheduling decision is made based on instantaneous observation in each time slot, and the trajectory decision is made based on average observation over multiple consecutive time slots. For T-UAV, it aims to optimize its user scheduling decisions and remain stationary due to its limited mobility capability. 
    
    \item We evaluate and validate the effectiveness of the proposed algorithm framework through extensive simulation results and comparison with benchmarks, including the two-timescale multi-agent proximal policy optimization (TTS-MAPPO) algorithm and MADDPG scheduling method. Our proposed TTS-MADDPG joint optimization method achieves a 12.2\% gain on the downlink successfully transmitted throughput compared to the MADDPG scheduling optimization method. The proposed algorithm also outperforms the TTS-MAPPO algorithm, with faster convergence, higher throughput, and stable performance. The effectiveness and good generalization capability of the proposed algorithm are further confirmed through the ablation study and parameter analysis.
\end{itemize}

The remainder of this paper is organized as follows: Section \ref{section_model_problem} presents the system model and problem formulation. Section \ref{section_problem_decomposition} provides the details of the problem decomposition in two timescales.
Section \ref{section_algorithm} details the proposed TTS-MADDPG algorithms. Section \ref{section_numerical_results} provides numerical results, including simulation settings, performance analysis, ablation study, and parameter analysis. Finally, Section \ref{section_conclusion} concludes the paper.

\section{System model and problem formulation}\label{section_model_problem}
In this section, we present our system model of a heterogeneous UAV-based cellular network for emergency communication in detail. This paper's main symbols and variables are listed in Table \ref{tab:symbols} for ease of reference.
\begin{table}[!t]
    \centering
    \renewcommand{\arraystretch}{1.2} 
    \caption{Table of Notations and Definitions}
    \label{tab:symbols}
    \begin{tabular}{|c|p{5.7cm}|}  
        \hline
        \textbf{Notations} & \textbf{Definition} \\
        \hline
        $k_0; \mathcal{K}_0; \mathcal{M}_{k_0}$ & T-UAV; T-UAV set; T-UAV associated G-UE set \\
        \hline
        $k_1; \mathcal{K}_1; \mathcal{M}_{k_1}$ & U-UAV; U-UAV set; U-UAV associated G-UE set \\
        \hline
        $m; \mathcal{M}$ & G-UE; Set of total G-UE \\
        \hline
        $\mathcal{K}$ & The set of total UAV \\
        \hline
        $T$ & Time slot length\\
        \hline        
        $A_t, A_u$ & The antenna number of T-UAV and U-UAV\\
        \hline        
        $\mathcal{C}^\mathrm{t}_\text{scd},\mathcal{C}^\mathrm{u}_\text{scd}$ & The maximum scheduling user number of T-UAV and U-UAV\\
        \hline
        $P_k$ & Transmission power of UAV $k$ \\
        \hline
        $B_k$ &  Bandwidth of UAV $k$ \\
        \hline
        $P_\text{intra}, P_\text{inter}$ & Power of intra-cell interference and inter-cell interference \\
        \hline
        $PL^\text{LoS}, PL^\text{NLoS}$ & LoS and NLoS path loss \\
        \hline
        $\mu_\text{LoS}, \mu_\text{NLoS}$ & LoS and NLoS attenuation factors \\
        \hline
        $T_\text{con}$ & Transmission buffer latency \\
        \hline
        $C_q$ & Quantized channel capacity \\
        \hline
        $N_\text{tx}$ & The number of successfully transmitted packets\\
        \hline
        $N_p$ & Packet size\\
        \hline
        $N_\text{str};N_\text{new};N_\text{cum}$ & Packets: Stored; Newly arrived; Accumulated before transmission; \\
        \hline
        $v_w; \bm{v}_{k_1}$ & Velocity of G-UE; Velocity of U-UAV \\
        \hline
        $\delta$ & Association status \\
        \hline
        $\gamma$ & Transmission Buffer status \\
        \hline
        $\zeta$ & Scheduling status \\
        \hline
        $\bm{g}; \bm{w}$ & Channel coefficient; Precoding vector \\
        \hline
        $\bm{\eta}$ & Rician fading coefficient  \\
        \hline
        $\tilde{K}$ & Rician factor \\
        \hline
        $\Theta$ & Elevation angle \\
        \hline
        $\phi_r, \phi_s$ & Angle of incidence of the LoS path on the receiver and transmitter antenna \\
        \hline
        $\bm{\pi};\bm{\mu}$ & Stochastic policy; Deterministic policy \\
        \hline
        $Q;\bar{Q}$ & Online Q-value; Target Q-value \\
        \hline
        $\beta$ & Discounting factor \\
        \hline
        $\bm{o},\bm{a},r$ & Partial observation, action, reward \\
        \hline
        $\mathbf{S},\mathbf{O},\mathbf{A},R$ & Global state, observation, action, reward \\
        \hline
        $\theta;\psi$ & Actor policy parameter ; Critic parameter \\
        \hline
        $\mathcal{J}$ & Policy objective function \\
        \hline
        $\mathcal{L}$ &  Critic loss function\\
        \hline
        $n;p$ & Short-timescale index; Long-timescale index \\
        \hline
    \end{tabular}
\end{table}

\begin{figure}[ht]
  \centering
  \includegraphics[width = 8.5cm]{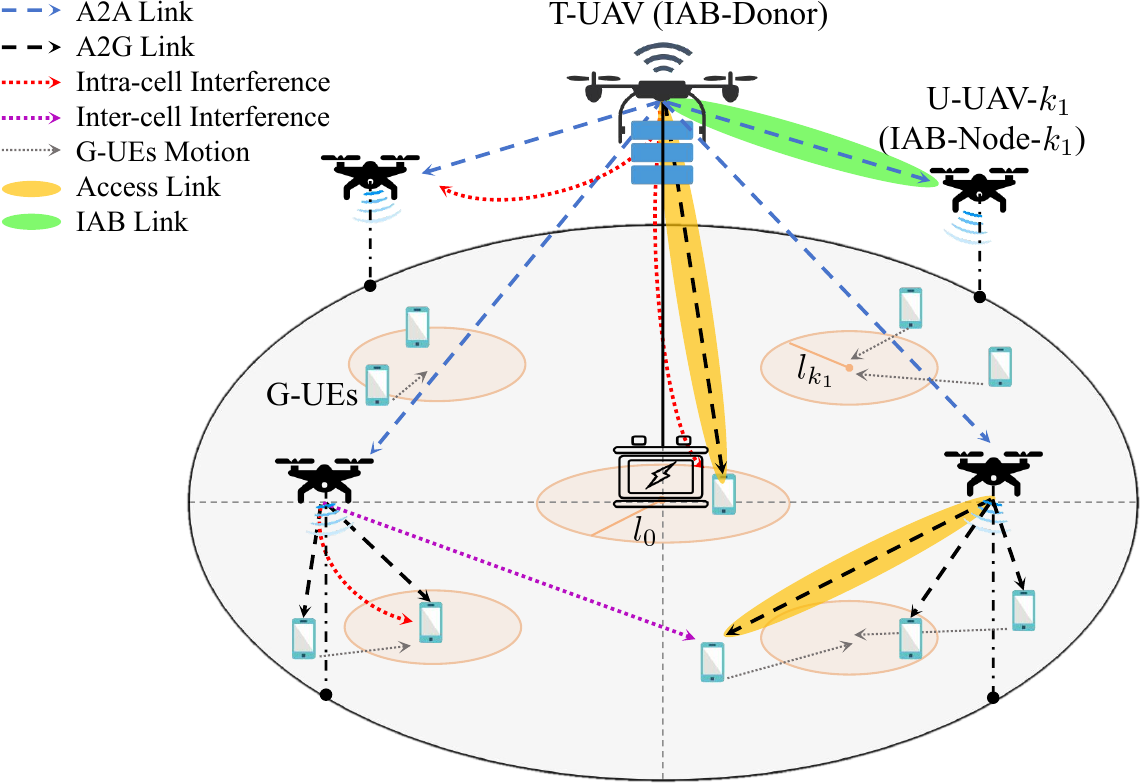}
  \caption{A typical system model of IAB-enabled heterogeneous UAV-based emergency communication network for post-disaster scenario.}
  \label{System_model}
\end{figure}

As shown in Fig. \ref{System_model}, we consider a circular geographic post-disaster area, where the UAVs are deployed to guarantee the time-critical downlink transmission for the G-UEs. There are totally $M$ G-UEs with a single antenna, denoted as $\mathcal{M} = \{1,\dots,m,\dots, M\}$, following the uniform distribution in the disaster area. We assume one T-UAV $k_0 \in\mathcal{K}_0 = \{0\}$ is located in the center of this disaster area and remains stationary with the height of $H_t$ because of its limited mobility caused by the tether. The T-UAV is equipped with $A_t$ antennas for access link communication with the associate G-UEs, denoted as $\mathcal{M}_{k_0} =\{ 1,2,..., M_{k_0}\}$, and IAB link communication with the U-UAVs, denoted as $\mathcal{K}_1 = \{1,2,..., K_1\}$ \cite{zhang2023packet,alavicheh2024integrated}. We assume each U-UAV $k_1$ serves the associated G-UEs $\mathcal{M}_{k_1} = \{1,2,..., M_{k_1}\}$ with $A_u$ antennas via access link communication \cite{selim2018post}. Each U-UAV has the same height of $H_u$. For convenience, we define the whole UAV group as $\mathcal{K}= \mathcal{K}_1 \cup \mathcal{K}_0$ and each UAV as $ k\in \mathcal{K}$. 
We assume the velocity of mobile G-UEs is $v_w$, and each U-UAV $k_1$ optimizes its trajectory by adjusting its velocity $\bm{v}_{k_1}[n] = \left[v^{k_1}_x[n],v^{k_1}_y[n]\right]$ to support the communication service to mobile G-UEs.
Without loss of generality, we assume each UAV adopts an equal power allocation scheme for its downlink transmissions among its scheduled users in each time slot, and the total transmission power of each UAV $k$ is denoted as $P_k$. We denote the bandwidth for UAV $k$ as $B_k$.

\subsection{Post-disaster Communication Phases}
To model the post-disaster emergency rescue, we assume a large circular safe zone is established in the center of the whole disaster area, providing shelters to the G-UEs connected to the T-UAV. Apart from that, four small circular safe zones are established in the center of each quadrant for the G-UEs associated with the U-UAV $k_1$. Without loss of generality, the emergency communication procedure can be mainly divided into two phases:
\begin{itemize}
    \item Initial Connection Phase: The T-UAV is deployed in the center of the whole disaster area for cell-center G-UEs, while the U-UAVs are deployed at the center of the edge of each quadrant for cell-edge G-UEs, and each G-UE is associated with only one UAV.
    \item Mobile G-UE Phase: Each G-UE moves toward its designated safe zone, where both the T-UAV and U-UAVs are required to ensure continuous and reliable communication services.
\end{itemize} 

\subsection{Channel Model}
We first model the geographical locations of both UAVs and G-UEs. We define the whole time duration as $T_w>0$ and divide it into $N$ equal time slot $T$, i.e., $T_w = T\cdot N$. 

The location of T-UAV $k_0$ is fixed and denoted as $L^u_{k_0} = \big(x_{k_0},y_{k_0},z_{k_0}\big) = \big(0,0,H_t\big)$. At time slot $n$, the location of the U-UAV $k_1$ is denoted as $L^u_{k_1}[n] = \big(x_{k_1}[n],y_{k_1}[n],z_{k_1}\big)$, and the location of G-UE $m$ is denoted as $L^g_{m}[n] = \big(x_{m}[n],y_{m}[n],0\big)$. The U-UAVs are initially deployed at the center of the edge of each quadrant, e.g., $L^u_{k_1}[1] = \big(\frac{\sqrt{2}}{2}l,\frac{\sqrt{2}}{2}l,H_u\big)$. Thus, the A2G distance between the UAV $k$ and G-UE $m$, and aerial-to-aerial (A2A) distance between T-UAV $k_0$ and U-UAV $k_1$ can be obtained as: 
\begin{equation}
    \label{space_distance_general}
    \begin{cases}
    d_{k,m}[n] = \|L^{u}_{k}[n]-L^{g}_{m}[n]\|_2, \\
    d_{k_0,k_1}[n] = \|L^{u}_{k_0}-L^{u}_{k_1}[n]\|_2.
    \end{cases} 
\end{equation}
We consider the full-duplex out-of-band IAB configuration, where the T-UAV and U-UAV operate on different frequency bands \cite{abdel2019uav}, with $f_t$ for T-UAV and $f_u$ for U-UAV. Moreover, we consider the uniform linear array (ULA) for both the T-UAV and U-UAVs, with half-wavelength array spacing $\lambda_{t}/2$ and $\lambda_{u}/2$, respectively.
Then, we model the A2G and A2A channels in our scenario as follows.

\subsubsection{A2G Channel}
As reported in practical experiments, the UAV at a sufficiently high altitude can establish line-of-sight (LoS) links with the G-UE and also experiences small-scale fading due to rich scattering \cite{khawaja2019survey}. Hence, the A2G link from UAV to G-UE consists of both LoS and non-line-of-sight (NLoS) components. We utilize the widely adopted probability path loss model for UAV communication as: 
\begin{equation}
    \label{path_loss_case}
    PL_{k,m}[n] \!=\! 
    \begin{cases}
    \left(\dfrac{4\pi d_{k,m}[n]}{\lambda_{t/u}}\right)^\alpha \!\mu_{\text{LoS}}, \!\!\!& \text{$Pr^{\text{LoS}}_{k,m}[n]$} \\
    \left(\dfrac{4\pi d_{k,m}[n]}{\lambda_{t/u}}\right)^\alpha \!\mu_{\text{NLoS}}, \!\!\!& \text{$Pr^{\text{NLoS}}_{k,m}\! =\! 1 \!-\! Pr^{\text{LoS}}_{k,m}[n]$,}
    \end{cases} 
\end{equation}
where $\alpha$ is the path loss exponent, $\lambda_{t/u}$ denotes the wavelength of the transmitted signal from T-UAV or U-UAV, $\mu_{\text{LoS}}$ and $\mu_{\text{NLoS}}$ are the attenuation factors for LoS and NLoS, and $Pr^{\text{LoS}}_{k,m}[n]$ is the probability of LoS, calculated by
\begin{equation}
    \label{los_probability}
    Pr^{\text{LoS}}_{k,m}[n] \approx \frac{1}{1+a\exp{\big(-b(\Theta_{k,m}[n]-a)\big)}},
\end{equation}
where $\Theta_{k,m}[n] = \frac{180^\circ}{\pi}\arcsin{\Big(\frac{z_{k}}{d_{k,m}[n]}\Big)}$ is the elevation angle between UAV $k$ and G-UE $m$ at time slot $n$, $a$ and $b$ are positive constants that depends on the environment \cite{qiu2020multiple,su2023energy}.

Hence, based on \eqref{path_loss_case}\eqref{los_probability}, the A2G large-scale path loss is expressed as follows:
\begin{equation}
    \label{A2G_large_path_loss}
    h_{k,m}[n] = \Big(\frac{4\pi d_{k,m}[n]}{\lambda_{t/u}}\Big)^\alpha \big(\mu_{\text{LoS}}Pr^{\text{LoS}}_{k,m}[n]+\mu_{\text{NLoS}}Pr^{\text{NLoS}}_{k,m}[n]\big).
\end{equation}

Then, we model the multiple-input single-output (MISO) A2G channel small-scale Rician fading coefficient as:
\begin{equation}
    \label{A2G_rician_coefficient}
    \bm{\eta}_{k,m}[n] \!=\! \sqrt{\frac{\tilde{K}_{k,m}[n]}{\tilde{K}_{k,m}[n]+1}}\bm{\eta^\text{LoS}}_{k,m}[n] \!+\! \sqrt{\frac{1}{\tilde{K}_{k,m}[n]+1}}\bm{\eta^\text{NLoS}}_{k,m}[n].
\end{equation}
In this equation, $\tilde{K}_{k,m}$ represents the Rician factor obtained by the following expression \cite{you20193d}:
\begin{equation}
    \label{rician_factor}
    \tilde{K}_{k,m}[n] = A_1 \exp({A_2\cdot \Theta_{k,m}[n]}),
\end{equation}
where $A_1$ and $A_2$ are the constant coefficients depending on the specific environment. 
% \textcolor{green}{We can also indicate the values we use for the $A_1$ or $A_2$.}
The $\bm{\eta^\text{LoS}}_{k,m}[n]$ is the deterministic LoS channel component given by 
% $|\bm{\hat{\eta}_{k_0,m}}| = \bm{1}$
\begin{equation}
    \label{Rician_LoS}
    \bm{\eta^\text{LoS}}_{k,m}[n] = e^{ - \frac{j 2 \pi d_{k,m}[n]}{\lambda_{t/u}}}\cdot\mathbf{e_t}(\phi_t),
\end{equation}
\begin{equation}
    \label{steering_vector_t}
    \mathbf{e_t}(\phi_t)  = \big[1, e^{-j\pi \cos\phi_t[n]},\dots,e^{-j\pi(A_{t/u}-1)\cos\phi_t[n]} \big]^\text{T}.
\end{equation}
where $\phi_t[n]$ represents the angle of incidence of the LoS onto the transmit antenna array and is calculated by $\phi_t[n] = \frac{\pi}{2}-\Theta_{k,m}[n]$, $A_{t/u}$ represents the antenna number $A_t$ for T-UAV or $A_u$ for U-UAV. $\bm{\eta^\text{NLoS}}_{k_0,m}[n]$ denotes the random scattering component with each of its elements following a zero-mean unit-variance circularly symmetric complex Gaussian (CSCG) \cite{cui2024near}. 

Therefore, the MISO A2G channel coefficient can be obtained as:
\begin{equation}
    \label{A2G_channel_coefficient}
    \bm{g}_{k,m}[n] = \sqrt{|h_{k,m}[n]|}\,\bm{\eta}_{k,m}[n]\in \mathbb{C}^{1 \times A_{t/u}}.
\end{equation}

\subsubsection{A2A Channel}
Due to the lack of scatters in A2A link \cite{do2021joint}, based on \eqref{path_loss_case}, the A2A large-scale path loss $h_{k_0,k_1}$ is expressed as:
\begin{equation}
    \label{A2A_large_path_loss}
        h_{k_0,k_1}[n] = \Big(\frac{4\pi d_{k_0,k_1}[n]}{\lambda_t}\Big)^\alpha \mu_{\text{LoS}}.
\end{equation}

Then, we model the multiple-input multiple-output (MIMO) A2A channel small-scale Rician fading coefficient $\bm{\eta}_{k_0,k_1}$ as:
\begin{equation}
    \label{A2A_rician_coefficient}
    \begin{aligned}
        &\bm{\eta}_{k_0,k_1}[n] =\\& \left(\sqrt{\frac{\tilde{K}_{k_0,k_1}[n]}{\tilde{K}_{k_0,k_1}[n]+1}}\bm{\eta^\text{LoS}}_{k_0,k_1}[n] \!+\! \sqrt{\frac{1}{\tilde{K}_{k_0,k_1}[n]+1}}\bm{\eta^\text{NLoS}}_{k_0,k_1}[n]\right),
    \end{aligned}
\end{equation}
where $\tilde{K}_{k_0,k_1}$ represents the Rician factor obtained by \eqref{rician_factor}, and the LoS channel component $\bm{\eta^\text{LoS}}_{k_0,k_1}$ can be calculated as:
\begin{equation}
    \label{transmit_steering_matrix}
    \bm{\eta^\text{LoS}}_{k_0,k_1}[n]  = \exp \left( - \frac{j 2 \pi d_{k_0,k_1}[n]}{\lambda_t} \right) \mathbf{e_r}(\phi_r) \mathbf{e_t}(\phi_t)^H,
\end{equation}
with $\mathbf{e_r}(\phi_r)$ is expressed as:
\begin{equation}
    \label{transmit_steering_vector}
    \mathbf{e_r}(\phi_r) = \big[1, e^{-j\pi  \cos\phi_r[n]},\dots,e^{-j\pi(A_u-1)\cos\phi_r[n]} \big]^\text{T},
\end{equation}
where $\phi_r[n]$ represents the angle of incidence of the LoS onto the transmit antenna array and is calculated by $\phi_r[n] = \frac{\pi}{2} - \Theta_{k_0,k_1}[n]$. In addition, the NLoS component $\bm{\eta^\text{NLoS}}_{k_0,k_1}$ is the random scattering component with elements following zero-mean unit-variance CSCG.

Therefore, the MIMO A2A channel coefficient can be obtained as:
\begin{equation}
    \label{A2A_channel_coefficient}
    \bm{g}_{k_0,k_1}[n] \!=\! \sqrt{|h_{k_0,k_1}[n]|}\cdot \bm{\eta}_{k_0,k_1}[n] \in \mathbb{C}^{A_u \times A_t}.
\end{equation}

\subsection{Ground User Equipment Association}
To ensure the stable emergency communication and the fairness among the G-UEs, we assume that each G-UE is only associated with the UAV based on the strongest received signal strength indicator (RSSI) at the first time slot, and the association decision remains fixed throughout the post-disaster communication phase.

Based on A2G large-scale path loss in \eqref{A2G_large_path_loss}, we can obtain the RSSI between UAV $k$ and G-UE $m$ by: 
\begin{equation}
    \label{RSSI_formula}
    \text{RSSI}^k_m[n] = {P_k} \cdot {h_{k,m}[n]}.
\end{equation}
Then, for G-UE $m$, based on the RSSI at the first time slot, its association status is denoted as:
\begin{equation}
    \label{user_association}
    \delta_{k,m} = 
    \begin{cases} 
    1,  \quad\text{if } k = \underset{i \in \mathcal{K}}{\argmax} \, \text{RSSI}^i_m[1], \\
    0,  \quad\text{else}.
    \end{cases}
\end{equation}
where $\delta_{k,m} = 1$ represents the G-UE $m$ is associated with UAV $k$, otherwise, $\delta_{k,m}=0$. 

\subsection{Downlink Transmission Scheme}\label{Downlink_transmission}
Without loss of generality, we assume a block fading channel in our scenario, where the channel state information (CSI) remains constant within each time slot. For clarity and conciseness, the subsequent derivations focus on a typical time slot, and the time index $[n]$ is dropped in this subsection. 

To model the user scheduling decision of T-UAV and U-UAV, we denote the scheduling status by $\zeta_{k,m}\in \!\{0,1\}$ for A2G link, and $\zeta_{k_0,k_1}\in \!\{0,1\}$ for A2A link. It is worth noting that we assume T-UAV $k_0$ can schedule at most $\mathcal{C}^\mathrm{t}_\text{scd}$ users among both its associated G-UEs and the U-UAVs, i.e., $\mathcal{M}_{k_0}\cup \mathcal{K}_1$, while the U-UAV $k_1$ can schedule at most $\mathcal{C}^\mathrm{u}_\text{scd}$ users among its associated G-UEs, i.e., $\mathcal{M}_{k_1}$.

\subsubsection{T-UAV to G-UE transmission}
The A2G transmission between the T-UAV and G-UE is modeled as a MISO system, where the received signal at the G-UE $m \in \mathcal{M}_{k_0}$ is presented as:
% consists of the information signal, intra-cell interference signals sent to other G-UEs $(\mathcal{M}_{k_0}^{m}\!=\!\mathcal{M}_{k_0}\setminus \{m\})$ or the U-UAVs $(\mathcal{K}_1)$, and the noise. The received signal is given as follows.
\begin{equation}
    \label{received_signal_BS2GUE}
    y_{k_0,m} = \sqrt{P_{k_0,m}} \,\bm{g}_{k_0,m}\,\bm{w}_{k_0,m} \,x_{k_0,m}\, + \mathcal{I}^{k_0,m}_{\text{intra}} +  n_{k_0,m}\,,
\end{equation}
\begin{equation}
    \label{intracluster_interference_UAVBS2GUE}
    \begin{aligned}
    \mathcal{I}^{k_0,m}_{\text{intra}} =  
    &\underbrace{\underset{j \in \mathcal{M}_{k_0}^{m}}{\sum}\!\zeta_{k_0,j}\sqrt{P_{k_0,m}} \,\bm{g}_{k_0,m}\,\bm{w}_{k_0,j} \,x_{k_0,j}}_{\text{I}} \!\\
    &+ \underbrace{\underset{i \in \mathcal{K}_1}{\sum}\!\zeta_{k_0,i} \sqrt{P_{k_0,m}} \,\bm{g}_{k_0,m}\,\bm{w}_{k_0,i} \,x_{k_0,i}}_{\text{II}},
    \end{aligned}
\end{equation}
where $\text{I}$ represents the intra-cell interference from scheduling other G-UEs, i.e., ($\mathcal{M}_{k_0}^{m}\!=\!\mathcal{M}_{k_0}\setminus \{m\}$), and $\text{II}$ represents the intra-cell interference from scheduling U-UAVs. The $P_{k_0,m}$ is the transmit power of T-UAV $k_0$ allocated for G-UE $m$, $\bm{g}_{k_0,m}$ is the channel coefficient obtained by \eqref{A2G_channel_coefficient}, $\bm{w}_{k_0,m}$ is the precoding vector, $x_{k_0,m}$ is the information signal with power $\mathbb{E}\{|x_{k_0,m}|^2\} = 1$, and $n_{k_0,m} \!\sim\!\mathcal{CN}(0,\sigma_{k_0,m}^2)$ is the additive white Gaussian noise (AWGN). 

Based on the maximum ratio transmission (MRT) technique, the precoding vector is obtained as:
\begin{equation}
    \label{BS2GUE_precoder}
    \bm{w}_{k_0,m} = \frac{\bm{g}_{k_0,m}^H}{\|\bm{g}_{k_0,m}\|}.
\end{equation}
Therefore, the signal-to-interference-plus-noise ratio (SINR) can be formulated as:
\begin{equation}
    \label{SINR_BS2GUE}
    \text{SINR}_{k_0,m} = \frac{P_{k_0,m}\cdot|\bm{g}_{k_0,m}\bm{w}_{k_0,m}|^2 }{P^{k_0,m}_{\text{intra}} +\sigma_{k_0,m}^2},
\end{equation}
where the power of intra-cell interference is obtained as:
\begin{equation}
    \label{power_intra_interference_UAVBS2GUE}
    \begin{aligned}
    P^{k_0,m}_{\text{intra}} = 
    &\underset{j \in \mathcal{M}_{k_0}^{m}}{\sum}\!\zeta_{k_0,j}P_{k_0,m}\cdot| \,\bm{g}_{k_0,m}\,\bm{w}_{k_0,j}|^2  \\
    & + \underset{i \in \mathcal{K}_1}{\sum}\!\zeta_{k_0,i} P_{k_0,m}\cdot\| \,\bm{g}_{k_0,m}\,\bm{w}_{k_0,i}\|^2.
    \end{aligned}
\end{equation}

\subsubsection{T-UAV to U-UAV transmission}
The A2A transmission between T-UAV and U-UAV is modeled as a MIMO system, where the received signal at the U-UAV $k_1$ is presented as:
\begin{equation}
    \label{received_signal_BS2UAV}
    \bm{y}_{k_0,k_1} = \sqrt{P_{k_0,k_1}} \,\bm{g}_{k_0,k_1}\,\bm{w}_{k_0,k_1} \,\bm{x}_{k_0,k_1}\, + \bm{\mathcal{I}}^{k_0,k_1}_{\text{intra}} + \bm{n}_{k_0,k_1}\,,
\end{equation}
\begin{equation}
    \label{intracluster_interference_UAVBS2UAVUE}
    \begin{aligned}
    \bm{\mathcal{I}}^{k_0,k_1}_{\text{intra}} =  
    &\underbrace{\underset{i \in \mathcal{K}_1^{k_1}}{\sum}\!\zeta_{k_0,i}\sqrt{P_{k_0,k_1}} \,\bm{g}_{k_0,k_1}\,\bm{w}_{k_0,i} \,\bm{x}_{k_0,i}}_{\text{III}}\!\\
    &+\underbrace{\underset{j \in \mathcal{M}_0}{\sum} \!\zeta_{k_0,j}\sqrt{P_{k_0,k_1}} \,\bm{g}_{k_0,k_1}\,\bm{w}_{k_0,j} \,\bm{x}_{k_0,j}}_{\text{IV}},
    \end{aligned}
\end{equation}
where $\text{III}$ represents the intra-cell interference from scheduling other U-UAVs, i.e., $(\mathcal{K}_1^{k_1} = \mathcal{K}_1\!\setminus\!\{k_1\})$, and $\text{IV}$ represents the intra-cell interference from scheduling G-UEs. The $P_{k_0,k_1}$ is the transmit power of T-UAV $k_0$ allocated for U-UAV $k_1$, $\bm{g}_{k_0,m}$ is the channel coefficient obtained by \eqref{A2A_channel_coefficient}, $\bm{w}_{k_0,k_1}$ is the MRT vector defined by \eqref{BS2GUE_precoder}, $\bm{x}_{k_0,k_1} \in \mathbb{C}^{A_u \times 1}$ is the information vector with unit power, and $\bm{n}_{k_0,k_1} \!\sim\!\mathcal{CN}(0,\sigma_{k_0,k_1}^2)$ is AWGN. 

Therefore, the SINR can be formulated as:
\begin{equation}
    \label{SINR_BS2UAV}
    \text{SINR}_{k_0,k_1} = \frac{P_{k_0,k_1} \cdot\|\bm{g}_{k_0,k_1}\bm{w}_{k_0,k_1}\|^2 }{P^{k_0,k_1}_{\text{intra}} +\sigma_{k_0,k_1}^2},
\end{equation}
where the power of intra-cell interference is obtained as:
\begin{equation}
    \label{power_intra_interference_UAVBS2UAVUE}
    \begin{aligned}
    P^{k_0,k_1}_{\text{intra}} = 
    &\underset{i \in \mathcal{K}_1^{k_1}}{\sum}\!\zeta_{k_0,i}P_{k_0,k_1}\cdot \|\bm{g}_{k_0,k_1}\,\bm{w}_{k_0,i}\|^2 \\
    &+\underset{j \in \mathcal{M}_0}{\sum} \!\zeta_{k_0,j}P_{k_0,k_1}\cdot \|\bm{g}_{k_0,k_1}\,\bm{w}_{k_0,j}\|^2.
    \end{aligned}
\end{equation}

\subsubsection{U-UAV to G-UE transmission}
The A2G transmission between U-UAV and G-UE is also modeled as a MISO system, where the received signal at the G-UE $m\in \mathcal{M}_{k_1}$ is presented as:
% consists of the information signal, intra-cell interference $(\mathcal{M}_{k_1}^{m} = \mathcal{M}_{k_1}\!\setminus \!\{m\})$, inter-cell interference signals sent by other U-UAV $(\mathcal{K}_1^{k_1})$, and the noise. Therefore, the received signal is expressed by 
\begin{equation}
    \label{received_signal_UE2GUE}
    y_{k_1,m} = \sqrt{P_{k_1,m}} \,\bm{g}_{k_1,m}\,\bm{w}_{k_1,m} \,x_{k_1,m}\, + \mathcal{I}^{k_1,m}_{\text{intra}} + \mathcal{I}^{k_1,m}_{\text{inter}} + n_{k_1,m}\,,
\end{equation}
\begin{equation}
    \label{intracluster_interference_UAVUE2GUE}
    \mathcal{I}^{k_1,m}_{\text{intra}} =  \underset{j \in \mathcal{M}_{k_1}^{m}}{\sum} \zeta_{k_1,j} \!\sqrt{P_{k_1,m}} \,\bm{g}_{k_1,m}\,\bm{w}_{k_1,j} \,x_{k_1,j},
\end{equation}
\begin{equation}
    \label{intercluster_interference}
    \mathcal{I}^{k_1,m}_{\text{inter}} = \underset{i \in \mathcal{K}_1^{k_1}}{\sum}\underset{j \in \mathcal{M}_{i}}{\sum} \!\zeta_{i,j}\sqrt{P_{i,j}} \,\bm{g}_{i,m}\,\bm{w}_{i,j} \,x_{i,j},
\end{equation}
where $\mathcal{I}^{k_1,m}_{\text{intra}}$ represents the intra-cell interference from scheduling other G-UEs, and $\mathcal{I}^{k_1,m}_{\text{inter}}$ represents the inter-cell interference from other U-UAVs' transmission signals. The $P_{k_1,m}$ is the transmit power of U-UAV $k_1$ allocated for G-UE $m$, $\bm{g}_{k_1,m}$ is the channel coefficient obtained by \eqref{A2G_channel_coefficient}, $\bm{w}_{k_1,m}$ is the MRT precoding vector defined by \eqref{BS2GUE_precoder}, $x_{k_1,m} \in \mathbb{C}$ is the information signal with unit power, and $n_{k_1,m} \sim\mathcal{CN}(0,\sigma_{k_1,m}^2)$.

Therefore, the SINR can be formulated as:
\begin{equation}
    \label{SINR_UAV2GUE}
    \text{SINR}_{k_1,m} = \frac{P_{k_1,m}\cdot|\bm{g}_{k_1,m}\bm{w}_{k_1,m}|^2 }{ P^{k_1,m}_{\text{intra}} + P^{k_1,m}_{\text{inter}} +\sigma_{k_1,m}^2},
\end{equation}
where the power of intra-cell and inter-cell interference can be obtained as:
\begin{equation}
    \label{power_intracluster_interference_UAVUE2GUE}
    P^{k_1,m}_{\text{intra}} =  \underset{j \in \mathcal{M}_{k_1}^{m}}{\sum} \zeta_{k_1,j} P_{k_1,m}\cdot |\bm{g}_{k_1,m}\bm{w}_{k_1,j}|^2 ,
\end{equation}
\begin{equation}
    \label{power_intercluster_interference}
    P^{k_1,m}_{\text{inter}} = \underset{i \in \mathcal{K}_1^{k_1}}{\sum}\underset{j \in \mathcal{M}_{i}}{\sum} \zeta_{i,j}P_{i,j}\cdot| \bm{g}_{i,m}\bm{w}_{i,j}|^2.
\end{equation}

\subsection{Traffic Management}
We consider the downlink burst traffic model in the scenario, where the number of newly arrived packets for each G-UE $m$ is modeled as an identically independent Poisson process with $ N^m_{\text{new}} \sim \text{Poisson}(\lambda) $. Without loss of generality, we assume each U-UAV maintains a first-in first-out (FIFO) buffer for its associated G-UEs, while the T-UAV maintains a local FIFO buffer for all G-UEs due to its connections to the core network. We consider that each packet has $N_p$ bits with a latency constraint $T_\text{con} = N_\text{con}\cdot T$, indicating that the packet will be dropped when exceeding this latency. 
\begin{figure}[ht]
  \centering
  \includegraphics[width = 8cm]{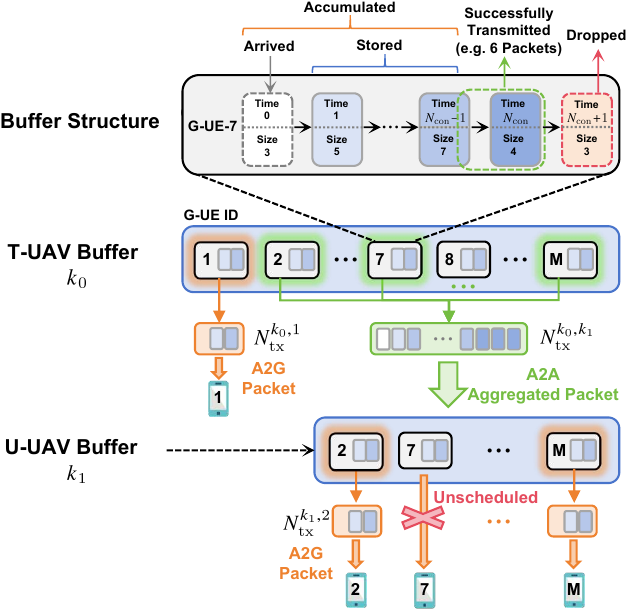}
  \caption{Traffic management process including both A2G and A2A transmissions.}
  \label{Transmission_Buffer}
\end{figure}

We assume UAV only schedules the associated G-UEs whose buffer in the corresponding UAV is not empty. T-UAV only schedules the U-UAV whose associated G-UEs' buffers in T-UAV are not all empty. Therefore, given the transmitter $i$ and receiver $j$, with $\{i\in\mathcal{K},j\in\mathcal{M}\}$ for A2G link or $\{i = k_0,j\in\mathcal{K}_1\}$ for A2A link, we define the non-empty buffer indication $\gamma_{i,j}[n] \in \{0,1\}$ as:
\begin{equation}
    \label{Buffer_status}
    \gamma_{i,j}[n] = \mathbbm{1}\big\{{N_{\text{cum}}^{i,j}[n]>0}\big\},
\end{equation}
where $\gamma_{i,j}[n] = 1$ indicates that the buffer in transmitter $i$ for receiver $j$ is not empty.  The $\mathbbm{1}\{\}$ is the indicator function that takes the value 1 if the statement $\mathbbm{1}\{\cdot\}$ is true, and zero otherwise. Based on Fig. \ref{Transmission_Buffer}, we use $N_{\text{cum}}^{i,j}[n]$ to represent the accumulated packets of the buffer in transmitter $i$ for receiver $j$ before transmission at time slot $n$, which is given by:
\begin{equation}
    \label{accumulated_packets}
    N_{\text{cum}}^{i,j}[n] = N_{\text{new}}^{i,j}[n] + N_{\text{str}}^{i,j}[n-\!1],
\end{equation}
where $N_{\text{new}}^{i,j}[n]$ denotes the newly arrived packets of the buffer in transmitter $i$ for receiver $j$ at time slot $n$, and $N_{\text{str}}^{i,j}[n-1]$ denotes the stored unsent packets of the buffer in transmitter $i$ for receiver $j$ at before time slot $n\!-\!1$. 

We illustrate the traffic management process in the buffers of both T-UAV and U-UAV in Fig. \ref{Transmission_Buffer}, where the FIFO-based buffer structure, A2G and A2A transmission processes, and the scheduling status are included. We denote the number of transmitted packets to receiver $j$ from the buffer in transmitter $i$ at time slot $n$ as $N_\text{tx}^{i,j}[n]$. We also provide an example for T-UAV's A2G transmission $N_\text{tx}^{k_0,1}$, U-UAV's A2G transmission $N_\text{tx}^{k_1,2}$, and T-UAV's A2A transmission $N_\text{tx}^{k_0,k_1}$ in Fig. \ref{Transmission_Buffer}. Typically, we can calculate $N_\text{tx}^{i,j}[n]$ by
\begin{equation}
    \label{data_rate_direct}
    N_\text{tx}^{i,j}[n] = \min\Big\{C^{i,j}_q[n],N_{\text{cum}}^{i,j}[n]\Big\},
\end{equation}
where $C_q^{i,j}$ is the quantized channel capacity, which is obtained as:
\begin{equation}
    \label{Quantized_channel_capacity}
    C_q^{i,j}[n] = \Big\lfloor\frac{B_i\cdot\log_2(1+\text{SINR}_{i,j}[n])\cdot T}{N_p}\Big\rfloor.
\end{equation}

The number of A2G transmitted packets $N_\text{tx}^{k,m}[n]$ and the A2A transmitted packets $N_\text{tx}^{k_0,k_1}[n]$ can be directly obtained by \eqref{data_rate_direct}. Since the A2A transmitted packets from T-UAV to U-UAV are subsequently forwarded to U-UAV's associated G-UEs, they need to aggregate the packets intended for all G-UEs served by U-UAV, i.e., $\mathcal{M}_{k_1}$, as shown in \eqref{data_rate_indirect}.
\begin{equation}
    \label{data_rate_indirect}
    N_\text{tx}^{k_0,k_1}[n] = \mathlarger{\sum}\limits_{m \in\mathcal{M}_{k_1}}N_\text{A2A}^{k_0,m}[n],
\end{equation}
where $N_\text{A2A}^{k_0,m}$ represents the the number of transmitted packets for each G-UE $m$ associated with U-UAV $k_1$. To obtain the value of $N_\text{A2A}^{k_0,m}$, we first sort $N^{k_0,m}_{\text{cum}}$ packets for all G-UEs $(m \in \mathcal{M}_{k_1})$ in descending latency order, and assign one packet to each G-UE in a Round-robin manner. Then, we repeat this process from the G-UE with the highest latency until the $N_\text{tx}^{k_0,k_1}[n]$ calculated by \eqref{data_rate_direct} is guaranteed or the buffers for all the G-UEs $\mathcal{M}_{k_1}$ are empty.

Therefore, based on \eqref{accumulated_packets}\eqref{data_rate_direct}\eqref{data_rate_indirect}, we can obtain $N_{\text{new}}^{i,j}[n]$ as follows:
\begin{equation}
    \label{new_packets}
    \begin{cases}
    N_{\text{new}}^{k_0,k_1}[n]=\mathlarger{\sum}\limits_{z \in\mathcal{M}_{k_1}}N_{\text{new}}^{z}[n],  \quad \hfill\text{when } i = k_0, j = k_1, \\ 
    N_{\text{new}}^{k_0,m}[n]= N_{\text{new}}^{m}[n],  \quad\hfill\text{when } i = k_0, j = m\in\mathcal{M}, \vspace{5pt}\\
    N_{\text{new}}^{k_1,m}[n]= N_\text{A2A}^{k_0,m}[n],  \quad\hfill\text{when } i = k_1, j = m\in \mathcal{M}_{k_1}. \\
    \end{cases}
\end{equation}

\subsection{Mobility Model of G-UEs}
In this scenario, each G-UE $m$ moves toward its designated safe zone with velocity $v_w$. For G-UE $m$, at the first time slot, the initial location is denoted as $L^g_m[0] = [x_m[0],y_m[0],0]$, and the final destination of G-UE $m$ is randomly selected within its designated safe zone and then remains unchanged, which is denoted as $\hat{L}^g_m = [\hat{x}_m,\hat{y}_m,0]$. 
At any time slot $n$, we can denote the distance between the current location with the initial location as:
\begin{equation}
    \label{distance_GUE_calculation}
    \hat{d}^\text{}_m[n] = \min\left(v_w n,\,\,\hat{d}^{\max}_m\right),
\end{equation}
where $\hat{d}^{\max}_m = \|\hat{L}^{g}_{m}-L^{g}_{m}[0]\|_2$. Each coordinate element can be represented by:
\begin{equation}
    \label{mobility_GUE}
    \begin{cases}
    x_m[n] = x_m[0] + \hat{d}_m[n]\cos\xi_m, \\
    y_m[n] = y_m[0] + \hat{d}_m[n]\sin\xi_m,
    \end{cases} 
\end{equation}
where $\xi_m = \arctan \frac{\hat{y}_m-y_m[0]}{\hat{x}_m-x_m[0]}$. Therefore, the location of G-UE $m$ at time slot $n$ can be determined as $L^g_m[n] = \left[x_m[n],y_m[n],0\right]$.

\subsection{Problem Formulation}
In this subsection, we formulate the problem to optimize the scheduling decision matrices $\bm{\zeta}_{k}$, with $\bm{\zeta}_{k_0}\in\mathbb{R}^{N\times (K_1+M_{k_0})}$ for T-UAV and $\bm{\zeta}_{k_1}\in\mathbb{R}^{N\times M_{k_1}}$ for U-UAV $k_1$, and the velocity matrices $\bm{V}_{k_1} \in \mathbb{R}^{N\times 2}$, with each column as $\bm{v}_{k_1}[n]$. Our objective is to maximize the long-term downlink throughput in this emergency communication scenario.

Therefore, the optimization problem is formulated as:
\begin{equation}\label{problem_formulation}
    \begin{aligned}
        &\underset{\bm{V}_{k_1},\bm{\zeta}_k}{\textbf{maximize}}\,\,\underset{k \in \mathcal{K}}{\mathlarger{\sum}} \underset{m \in \mathcal{M}_k}{\mathlarger{\sum}} \overset{N}{\underset{n=1}{\mathlarger{\sum}}}\,\,\mathbb{E}\Big\{N_\text{tx}^{k,m}[n]\zeta_{k,m}[n]\Big\} \\
    \end{aligned}
\end{equation}
\begin{align*} 
% ---------- Velocity constraint $<$ v ----------- %
\,\,\,\,\textbf{subject to} \,\, &~|v_x^{k_1}|\leq v_d^{\max}, |v_y^{k_1}|\leq v_d^{\max}, 
\tag{40a}\label{CON_a}
% ---------- Antenna Number Constraint for T-UAV + U-UAV ------ %
\\&~\underset{m \in \mathcal{M}_0}{{\sum}}\underset{k_1 \in \mathcal{K}_1}{{\sum}}\Big(\zeta_{k_0,m}[n]+\zeta_{k_0,k_1}[n]\Big) \leq \mathcal{C}^\mathrm{t}_\text{scd}, \forall n,
\tag{40b}\label{CON_b}
\\&~\underset{m \in \mathcal{M}_{k_1}}{{\sum}}\zeta_{{k_1},m}[n] \leq \mathcal{C}^\mathrm{u}_\text{scd}, \forall n, 
\tag{40c}\label{CON_c}
\end{align*}
where \eqref{CON_a} represents the dimensional velocity constraint of U-UAVs, and \eqref{CON_b} and \eqref{CON_c} represent the maximum scheduling user number limits.

The downlink throughput optimization problem formulated above is a MINLP problem. Since the classical NP-complete problem, such as the 0--1 Knapsack problem \cite{chapman2001system}, is reducible to the MINLP problem, our optimization problem is also NP-hard \cite{belotti2013mixed,huang2019hybrid}. This problem is thus very difficult to solve in polynomial time by conventional optimization techniques, such as simplex or interior-point methods. This problem becomes even more complex to capture the real-time decision-making mechanism since it involves the mobility of UAVs, traffic arrival, and channel randomness. Specifically, in the absence of prior knowledge about the dynamic channel conditions and the network environment, traditional offline algorithms struggle with rendering real-time decisions to arrive at a solution for the problem. This is because the typical offline optimization algorithm needs to know all the state information of the network before solving the optimization problem. Therefore, traditional iterative offline algorithms make it hard to solve the problem timely. As a machine learning method, MADRL is capable of interacting and learning from the environment and finally obtains a policy model that can be deployed on the devices, thereby facilitating real-time decisions and meeting long-term benefits according to the current state.

\section{POMDP Formulation and Problem Decomposition }\label{section_problem_decomposition}
In this section, we introduce the partially observable Markov decision process (POMDP) and further decompose our formulated problems into two timescales with detailed observation, action, and reward settings.

\subsection{POMDP Formulation}
Traditional MDP-based optimization methods typically assume complete global observations to every agent, making them ineffective in dynamic and uncertain environments with multi-agent settings \cite{liu2024two}. Therefore, we formulate the problem as a POMDP to enable sequential decision-making under partial observability that can effectively handle environmental changes and inherent uncertainties. Generally, a POMDP of an agent set $\mathcal{K}$ can be generally denoted as $<\mathcal{K},\mathcal{O},\mathcal{S},\mathcal{A},\mathcal{P},\mathcal{R},\bm{\pi}>$, which is composed of observation space $\mathcal{O}$, state space $\mathcal{S}$, action space $\mathcal{A}$, probability of environment transferring $\mathcal{P}$, reward function $\mathcal{R}$, and stochastic policy $\bm{\pi}$. The $\bm{\pi}_{k}\left(a_{k}|o_{k}\right)$ denotes the probability of taking action $a_{k}$ at observation $o_{k}$. The deterministic policy is usually denoted as $\bm{\mu}_{k}(o_{k})$, which maps each observation directly to a specific action. In multi-agent settings, the global state $\mathbf{S} \in \mathcal{S}$ is partially observable to agents. Consequently, the agent $k$ can only get the partial observation $o_{k} \in \mathcal{O}$ from the environment. The immediate reward $r_k$ for each agent is obtained by the function $\mathcal{R}$ after the action $a_k$.

\subsection{Problem Decomposition}
The user scheduling decision is performed at the short timescale (every time slot) following 5G standards, which depends on the numerology of the 5G system (e.g., slot length). The UAV control and command (C\&C) signal is executed at the long timescale due to hardware limitations \cite{licea2024robotics}. For example, the control signal transmission of the DJI UAV should be larger than 40 ms \cite{zhou2021real}. This fact indicates that the trajectory control task should be executed at a relatively longer timescale compared to user scheduling decisions \cite{xu2023air}.
Therefore, considering the asynchronous update between user scheduling and trajectory control, we decompose the formulated optimization in \eqref{problem_formulation} into two timescales as follows:
\begin{equation}\label{decompose_scheduling}
    \begin{aligned}
        \text{Short-Timescale:}&\underset{\{\bm{\mu}^\text{S}(\bm{a}^\text{S}|\bm{o}^\text{S})\}}{\textbf{max}}\,\,\overset{N}{\underset{n=1}{\mathlarger{\sum}}} \beta^{n - 1}\mathbb{E}_{\bm{\mu}^\text{S}}\{R^\text{S}[n]\} \\
        & \quad \textbf{\textit{s. t. }} \,\, \eqref{CON_b}, \eqref{CON_c} 
    \end{aligned}
\end{equation}
where $\bm{\mu}^\text{S}$ is the deterministic policy that maps the current observation $\bm{o}^\text{S}$ to action $\bm{a}^\text{S}$, $n$ is the short-timescale index for the user scheduling update process, $\beta \in (0,1]$ is the discounting factor for the performance in future time slots, and $R^\text{S}$ is the short-timescale reward over all agents, which will be introduced later.
\begin{equation}\label{decompose_trajectory}
    \begin{aligned}
        \text{Long-Timescale:}&\underset{\{\bm{\mu}^\text{T}(\bm{a}^\text{T}|\bm{o}^\text{T})\}}{\textbf{max}}\,\,\overset{\lfloor{N/N_l}\rfloor}{\underset{p=1}{\mathlarger{\sum}}} \,\,\beta^{p - 1}\mathbb{E}_{\bm{\mu}^\text{T}}\{R^\text{T}[p]\} \\
        & \quad \textbf{\textit{s. t. }} \,\, \eqref{CON_a}, \eqref{CON_b}, \eqref{CON_c},
    \end{aligned}
\end{equation}
where $p$ is the long-timescale index for the trajectory control update process, and we assume the long-timescale length is $N_p$ times longer than the short-timescale index, i.e., $p = \lfloor\frac{n}{N_l}\rfloor$. The remaining parameters are defined similarly to the short-timescale parameters.

\subsection{User Scheduling Problem in Short Timescale}
This subproblem focuses on optimizing user scheduling decisions of each UAV to maximize the overall downlink successfully transmitted throughput, where each UAV is modeled as an agent responsible for determining its optimal scheduling strategy. The details about the agents are provided below. For clarity, unless otherwise specified, the term \textit{time slot} mentioned in this paper refers to the short-timescale time slot.

\begin{itemize}
    \item \textbf{Observation space $\mathcal{O}^\text{S}$}: At each time slot, UAV $k$ can observe its transmission buffer feature $\bm{b}_{k}$, the historical SINR of associated users $\text{SINR}_k$, the reward of previous time slot $r^{\text{S}^-}_k$, and the scheduling action of previous time slot $\bm{a}^{\text{S}^-}_k$. Hence, the observation is defined by 
    \begin{equation}
        \label{Observation_scheduling}
        \bm{o}^\text{S}_k = \{\bm{b}_k,\text{SINR}_k,r_k^{\text{S}^-},\bm{a}_k^{\text{S}^-}\}.
    \end{equation}
    Specifically, for U-UAV $k_1$ or T-UAV $k_0$, the transmission buffer features are given as:
    \begin{equation}
        \label{buffer_feature_uav}
        \begin{cases} 
        \bm{b}_{k_1} = &\big\{(N^{k_1,m}_\text{cum}, \Bar{T}^{k_1}_m, \hat{T}^{k_1}_m)\,|\,m \in \mathcal{M}_{k_1}\big\},  \hfill\hfill  \\ 
        \bm{b}_{k_0} = &\Big\{\big\{(N^{k_0,k_1}_\text{cum}, \Bar{T}^{k_0}_{k_1}, \hat{T}^{k_0}_{k_1})\,|\,k_1 \in \mathcal{K}_1\} \bigcup\\& \{(N^{k_0,m}_\text{cum}, \Bar{T}^{k_0}_m, \hat{T}^{k_0}_m)\,|\,m \in \mathcal{M}\big\}\Big\},
        \end{cases}
    \end{equation}
    where $\Bar{T}^{k_1}_m$ (or $\Bar{T}^{k_0}_m$) denotes the average queueing delay of packets and $\hat{T}^{k_1}_m$ (or $\hat{T}^{k_0}_m$) denotes the latency of the currently first packet for target G-UE $m$ in the buffer at UAV $k_1$ (or $k_0$). Meanwhile, $\Bar{T}^{k_0}_{k_1}$ and $\hat{T}^{k_0}_{k_1}$ are similarly defined but based on T-UAV's buffer for U-UAV $k_1$'s total G-UEs.
    
    \item \textbf{Action space $\mathcal{A}^\text{S}$}:
    The action of each agent is defined as the union of scheduling status towards its associated users. Specifically, for T-UAV $k_0$ and U-UAV $k_1$, their actions are expressed as:
    \begin{equation}
        \label{action_scheduling}
        \begin{cases}
            \bm{a}^\text{S}_{k_0} \!= \bm{\zeta}_{k_0} \!= \big\{\zeta_{k_0,j}|j\in(\mathcal{K}_1\!\cup\!\mathcal{M}_{k_0})\big\}\in \mathbb{R}^{1 \times (K_1+M_{k_0})}, \\
            \bm{a}^\text{S}_{k_1} \!= \bm{\zeta}_{k_1} \!= \big\{\zeta_{k_1,j}|j\in\mathcal{M}_{k_1}\big\}\in \mathbb{R}^{1 \times M_{k_1}}. \\
        \end{cases}
    \end{equation}
    To represent the joint user scheduling actions, we define:
    \begin{equation}
        \label{Joint_us_actions}
        \mathbf{A}^\text{S} = \{\bm{a}^\text{S}_0,\dots,\bm{a}^\text{S}_{K_1}\}.
    \end{equation}
    
    \item \textbf{Reward $\mathcal{R}^\text{S}$}: The immediate reward of agent $k$ is denoted by $r^\text{S}_k[n]$, which is the number of successfully transmitted packets at the current time slot $n$. 
    Its formula is given by
    \begin{equation}
        \label{reward_scd}
         r_k^\text{S}[n]= \underset{m\in\mathcal{M}_k}{\sum}N^{k,m}_\text{tx}[n]\zeta_{k,m}[n].
    \end{equation}
    In addition, we define $\mathbf{R}^\text{S}$ to represent the rewards for each agent, which is shown as $\mathbf{R}^\text{S} = \{r_0^\text{S},r_1^\text{S},\dots,r_{K_1}^\text{S}\}$. The global reward of all agents is given by 
    \begin{equation}
        \label{global_reward_scd}
         R^\text{S}[n] = \underset{k \in \mathcal{K}}{\sum}r_k^\text{S}[n].
    \end{equation}

    \item \textbf{State}: The global state is defined as the combination of all agents' partial observations, which is given by $\mathbf{S}^\text{S} = (\bm{o}^\text{S}_0,\dots, \bm{o}^\text{S}_{K_1})$. 
\end{itemize}
\begin{figure*}[!ht]
  \centering
  \includegraphics[width=\textwidth]{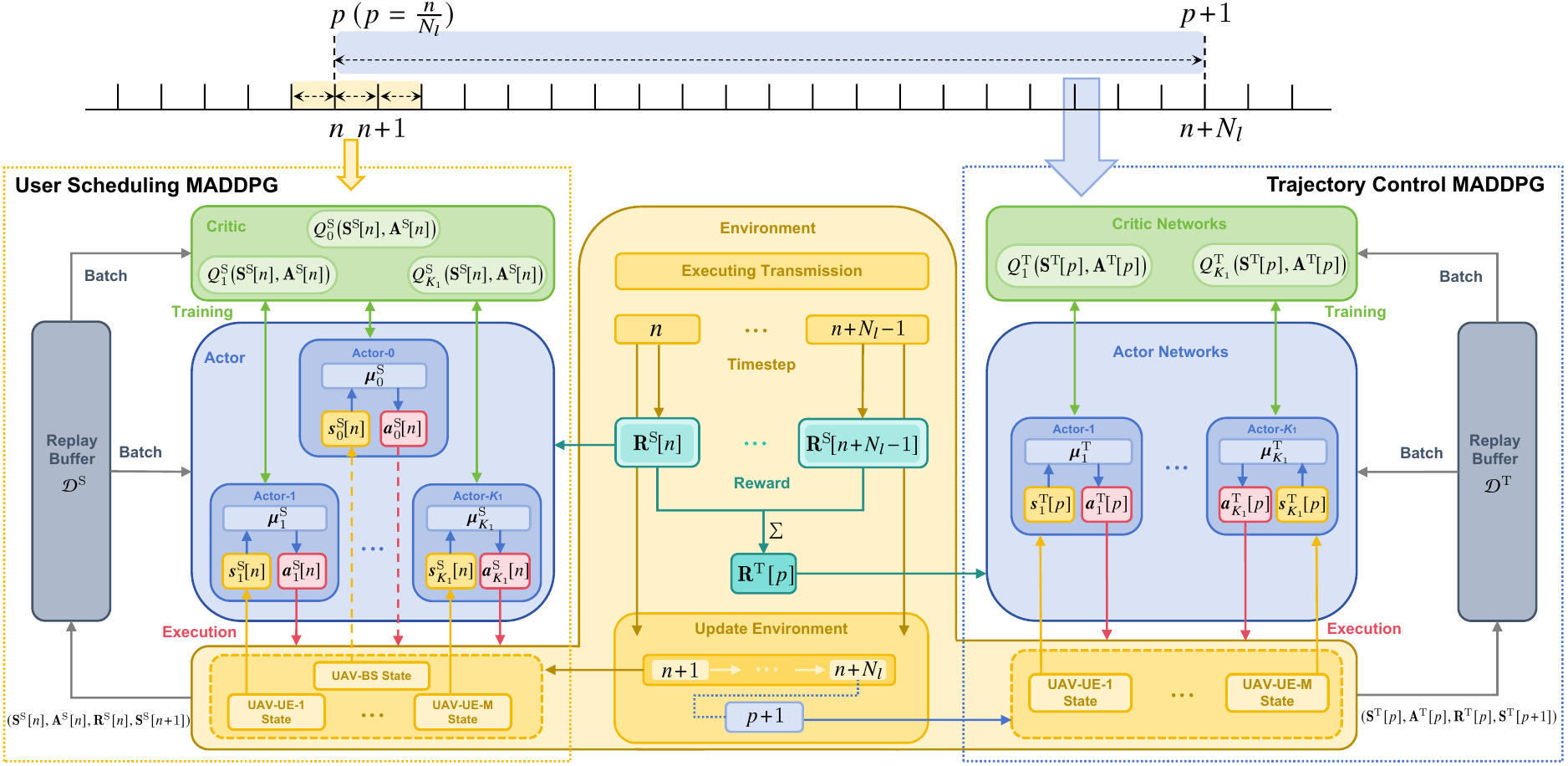}
  \caption{Overall structure and workflow of the proposed TTS-MADDPG algorithm.}
  \label{Algorithm_structure}
\end{figure*}

\subsection{UAV Trajectory Control Problem in Long Timescale}
This subproblem aims to design the optimal trajectory control strategy for U-UAVs to maximize the downlink successfully transmitted throughput, where each U-UAV is modeled as an agent responsible for determining its trajectory control actions. As the trajectory control actions are taken every long-timescale time slot $p = \lfloor\frac{n}{N_l}\rfloor$, the details about this subproblem are presented below.

\begin{itemize}
    \item \textbf{Observation space $\mathcal{O}^\text{T}$}: At each long-timescale trajectory update time slot $p$, each agent $k_1$ observes the RSSI for its associated G-UEs $\text{RSSI}_{k_1}$, its trajectory control action and reward of previous long-timescale time slot, $\bm{a}^{\text{T}^-}_{k_1}$ and $r^{\text{T}^-}_{k_1}$, the position of itself $L^u_{k_1}$, and the average positions of its associated G-UEs $\bar{L}^g_{m_{k_1}}$. Therefore, the local observation is given as
    \begin{equation}
        \label{observation_trajectory}
        \bm{o}_{k_1} = \{\text{RSSI}_{k_1}, r^{\text{T}^-}_{k_1}, \bm{a}^{\text{T}^-}_{k_1}, L^u_{k_1}, \bar{L}^g_{m_{k_1}}\}.
    \end{equation}
    
    \item \textbf{Action space $\mathcal{A}^\text{T}$}: Each agent outputs the action of velocity, expressed by 
    \begin{equation}
        \label{action_trajectory}
        \bm{a}^\text{T}_{k_1} = \bm{v}_{k_1} = \big[v^x_{k_1},v^y_{k_1}\big].
    \end{equation}
    The position change after action is $\Delta L^u_{k_1} = \bm{a}_{k_1}^\text{T}(N_l\cdot T)$. To represent the joint trajectory control actions, we define:
    \begin{equation}
        \label{Joint_tc_actions}
        \mathbf{A}^\text{T} = \{\bm{a}^\text{T}_1,\dots,\bm{a}^\text{T}_{K_1}\}.
    \end{equation}
    
    \item \textbf{Reward $\mathcal{R}^\text{T}$}: The long-timescale partial reward $r_{k_1}^\text{T}[p]$ is the summation of the past $N_l$ short-timescale rewards, which is given by: 
    \begin{equation}
        \label{reward_overall_scd}
         r_{k_1}^\text{T}[p]= \frac{1}{N_l}\sum^{n + N_l - 1}_{i = n}r^\text{S}_{k_1}[i].
    \end{equation}
    In addition, we define $\mathbf{R}^\text{T}$ to represent the rewards for each agent, which is shown as $\mathbf{R}^\text{T} = \{r_1^\text{T},\dots,r_{K_1}^\text{T}\}$. The long-timescale global reward of all agents is calculated by 
    \begin{equation}
        \label{global_reward_overall_scd}
         R^\text{T}[p] = \underset{k_1 \in \mathcal{K}_1}{\sum} r_{k_1}^\text{T}[p].
    \end{equation}
    
    \item \textbf{State}: The global state is defined as the combination of all agents' partial observations, which is given by $\mathbf{S}^\text{T} = (\bm{o}^\text{T}_1,\dots, \bm{o}^\text{T}_{K_1})$.

\end{itemize}

\section{The Proposed Algorithm}\label{section_algorithm}
In this section, we present our proposed hierarchical reinforcement learning algorithm to solve the above-formulated problem. First, the preliminaries related to DRL are presented. Then, we introduce the overall structure and workflow of our hierarchical TTS-MADDPG algorithm framework.

\subsection{Preliminaries of DRL}

Traditional RL algorithms, such as Q-Learning and DQN, have been widely applied in single-agent settings to solve sequential decision-making problems. These methods enable an agent to learn the stochastic policy $\bm{\pi}$ to maximize the expected discounted cumulative reward $\mathbb{E}_{\bm{\pi}}[G_t]$, defined as
\begin{equation}
    \label{Expected_discounted_reward}
    G_t = R_t + \beta R_{t+1} + \beta^2R_{t+2} + \dots = \overset{\infty}{\underset{i = 0}{\sum}}\beta^iR_{t+i},
\end{equation}
where $\beta$ is the discount factor. To find the optimal policy $\bm{\pi}^*$, the state action value function $Q(\mathbf{S},\mathbf{A})$, called Q-value, is introduced to estimate the expected discounted cumulative reward by executing an action $\mathbf{A}$ at state $\mathbf{S}$ under policy $\bm{\pi}$. The equation of the Q-value function based on the Bellman equation is given by
\begin{equation}
    \label{Q_value_intro}
    \begin{aligned}
        Q(\mathbf{S},\mathbf{A}) &= \mathbb{E}_{\bm{\pi}}[G_t|\mathbf{S}_t = \mathbf{S}, \mathbf{A}_t = \mathbf{A}],\\
        & = \mathbb{E}_{\bm{\pi}}[R_t + \beta Q(\mathbf{S}', \mathbf{A}')|\mathbf{S}_t = \mathbf{S}, \mathbf{A}_t = \mathbf{A}].
    \end{aligned}
\end{equation}

DQN utilizes deep neural networks to approximate the Q-value function in discrete action spaces \cite{guo2022multi}. However, DQN exhibits limitations when dealing with the optimization problem with continuous action \cite{arani2023haps}. To address this, the DDPG algorithm adopts an actor-critic architecture to learn a deterministic policy $\bm{\mu}_\theta$ and critic $Q$ based on deep neural networks, making it suitable for tasks such as trajectory optimization in our scenario. More importantly, for multi-agent scenarios, a CTDE-based algorithm called MADDPG is introduced \cite{lowe2017multi}, with each agent making decisions based on its partial observation instead of the global state. Motivated by this, we propose our algorithm based on MADDPG to solve the multi-agent optimization problems in our scenario.

\subsection{Proposed TTS-MADDPG Algorithm}
In this subsection, we introduce the overall structure and workflow of the hierarchical TTS-MADDPG algorithm for maximizing the downlink throughput.

Given the multi-agent scenario, the environment becomes non-stationary from any individual agent's perspective, resulting in an unstable learning process. Considering the necessary coordination among the UAV agents and the independent execution of each agent, this algorithm is designed based on the centralized training and distributed execution (CTDE) framework \cite{peng2025intelligent}. Specifically, the offline centralized training is usually implemented in a simulation environment, which avoids the challenges associated with high bandwidth overhead or latency. During centralized learning, global state information is utilized by the central critic to assist learning, while each agent’s actor is required to only access local observations. After completing the training, the distributed execution process is executed online. Each UAV independently executes its learned policy with its offline well-trained actor networks based on local observations, without relying on the global state information and the central critic. This makes the deployment feasible in practical UAV-enabled emergency communication scenarios, where global information exchange is costly or unavailable due to communication latency, limited bandwidth, etc.

What's more, each agent has neural networks with the actor-critic framework, which contains an individual actor network and a critic network. Each actor network or critic network consists of an online network and a target network, which have the same structure but different updated rate parameters. These target networks are established to make the online networks' learning process stable and convergent \cite{liu2024two,zhou2024joint}. Fig. \ref{Algorithm_structure} illustrates the algorithm structure, and the details of each part are introduced below.
\begin{algorithm}[ht]
    \caption{TTS-MADDPG Algorithm}
    \label{alg:two_timescale_maddpg}
    \begin{algorithmic}[1]
        \State Initialize actor networks $\bm{\mu}^\text{S}_i,\bm{\mu}^\text{T}_j$ with parameters $\theta^\text{S}_i, \theta^\text{T}_j$ , and critic networks $Q^\text{S},Q^\text{T}$ with parameters $\psi^\text{S}, \psi^\text{T}$.
        \State Initialize target actor network $\bar{\bm{\mu}}^\text{S}_i,\bar{\bm{\mu}}^\text{T}_j$ and target critic network $\bar{Q}^\text{S}_i,\bar{Q}^\text{T}_j$ with parameters $ \theta_i^{\text{S}'} \leftarrow \theta_i^\text{S} $, $ \theta_j^{\text{T}'} \leftarrow \theta_j^\text{T} $, $ \psi^{\text{S}'}_i \leftarrow \psi^\text{S}_i $, $ \psi^{\text{T}'}_j \leftarrow \psi^\text{T}_j $.
        \State Initialize replay buffer $ \mathcal{D}^\text{S} $ with batch size $B_\text{S}$, and $\mathcal{D}^\text{T}$ with batch size $B_\text{T}$.
        \State Initialize the episode length $L_e$, the maximum episodes $E$, the epsilon-greedy parameter $\epsilon$.
        \For{episode$ = 1,2,\dots,E$}
            \State Initialize environment and obtain initial state.
            \For{short-timescale step $ n = 1, 2,\dots, L_e$}
                \For{each agent $ i \in \mathcal{K}$}
                    \State Set action $ \bm{a}^\text{S}_i[n]$ based on $\epsilon$-greedy policy
                \EndFor
                \For{each agent $ j \in \mathcal{K}_1$}
                    \If{$ n \bmod N_l = 0 $}
                        \State Enter long-timescale step $p = n/N_l$
                        \State Set action $ \bm{a}^\text{T}_j[p]$ based on $\epsilon$-greedy policy
                    \EndIf
                \EndFor
                \State Execute joint action $ (\mathbf{A}^\text{S},\mathbf{A}^\text{T}) $, receive reward $ (\mathbf{R}^\text{S}, \mathbf{R}^\text{T})$ and next state $ (\mathbf{S}^{\text{S}'},\mathbf{S}^{\text{T}'})$.
                \State Store $ (\mathbf{S}^\text{S}, \mathbf{A}^\text{S}, \mathbf{R}^\text{S}, \mathbf{S}^{\text{S}'})$ in replay buffer $ \mathcal{D}^\text{S} $.
                \State Store $ (\mathbf{S}^\text{T}, \mathbf{A}^\text{T}, \mathbf{R}^\text{T}, \mathbf{S}^{\text{T}'})$ in replay buffer $ \mathcal{D}^\text{T} $.
                \If{buffer $ \mathcal{D}^\text{S} $ size $\geq$ $B_\text{S}$}
                    \For{each agent $ i \in \mathcal{K}$}
                        \State Sample batch $B_\text{S}$ from $ \mathcal{D}^\text{S} $.
                        \State Update online $\bm{\mu}^\text{S}_i$ and $Q^\text{S}_i$ by \eqref{actor_para_update}\eqref{critic_para_update}.
                        \State Update target $\bar{\bm{\mu}}^\text{S}_i$ and $\bar{Q}^\text{S}_i$ by \eqref{target_actor_critic_para_update}.
                    \EndFor
                \EndIf
                \If{$ n \bmod N_l = 0 $ and buffer $ \mathcal{D}^\text{T} $ size $\geq$ $B_\text{T}$}
                    \For{each agent $ j \in \mathcal{K}_1$}
                        \State Sample batch $B_\text{T}$ from $ \mathcal{D}^\text{T} $.
                        \State Update online $\bm{\mu}^\text{T}_j$ and $Q^\text{T}_j$ by \eqref{actor_para_update}\eqref{critic_para_update}.
                        \State Update target $\bar{\bm{\mu}}^\text{T}_j$ and $\bar{Q}^\text{T}_j$ by \eqref{target_actor_critic_para_update}.
                    \EndFor
                \EndIf
            \EndFor
        \EndFor
    \end{algorithmic}
\end{algorithm}

\textbf{Actor}: The actor network aims to approximate the optimal action policy and output the actions based on its partial observation. Two groups of agents are designed for different tasks: one group ($\mathcal{K}$) is responsible for user scheduling, while the other group ($\mathcal{K}_1$) focuses on trajectory control. The online and target actor networks employ deterministic policies, parameterized by \( \theta^\text{S} \) and \( \theta^{\text{S}'} \) for the user scheduling agent, and \( \theta^\text{T} \) and \( \theta^{\text{T}'}\) for the trajectory control agent. Each user scheduling actor executes the action at time slot ($n$), while each trajectory control actor executes the action at time slot ($p$). 
For clarity and conciseness, in the following introductions, we will omit the explicit notation for the user scheduling task (denoted by superscript $\text{S}$) and the trajectory control task (denoted by superscript $\text{T}$), as the formulas mentioned later can be applied to both tasks.

Generally, for agent $i$, we define $\bm{\mu}_i\big(\bm{o}_i|\theta_i\big)$  (abbreviated as $\bm{\mu}_i$) as the action policy functions. 
To find an optimal action policy that helps maximize the expected long-term cumulative reward $G_t$, the policy objective function is denoted as 
\begin{equation}
    \label{objective_function_policy}
    \mathcal{J}\big(\bm{\mu}_i \big) = \mathbb{E}_{\theta_i}\big[G_t \big], \quad \text{with}\,\,\, G_t = \sum^{\infty}_{j = 0}\beta^j R[t+j],
\end{equation}
where $t$ can be either a short-timescale or a long-timescale time slot, depending on the tasks.
Hence, the optimal action policy $\bm{\mu}^*$ will be obtained by exploring the corresponding parameters $\theta_i$ to maximize the objective functions, i.e., 
\begin{equation}
    \label{optimal_policy_US}
    \bm{\mu}^*_i= \underset{\theta_i}{\argmax}\,\,\mathcal{J}_{\bm{\mu}_i}\big(\theta_i \big),
\end{equation}

Furthermore, the gradient of these objective functions can be written as \eqref{actor_gradient}, used by both policies in the future gradient descent or ascent process.
\begin{equation}
    \label{actor_gradient}
    \begin{aligned}
        &\nabla_{\theta_i} \mathcal{J}(\bm{\mu}_i) \\
        & = \mathbb{E}_{\mathbf{S}, \mathbf{A} \sim \mathcal{D}} \Big[ \nabla_{\theta_i} \bm{\mu}_i (\bm{a}_i|\bm{o}_i) \nabla_{\bm{a}_i} Q_i (\mathbf{S}, \mathbf{A}) \big|_{\bm{a}_i = \bm{\mu}_i(\bm{o}_i)} \Big],
    \end{aligned}
\end{equation}
where $\mathcal{D}$ represents the replay buffer, which records the experiences of all agents in the form of a tuple $(\mathbf{S}, \mathbf{A}, \mathbf{R}, \mathbf{S}')$, shown in the grey blocks in Fig. \ref{Algorithm_structure}. During training, based on a batch of sampled experiences from $\mathcal{D}$, the gradient is back-propagated to the online actor network to update $\theta_i$ by
\begin{equation}
    \label{actor_para_update}
    \theta_i \leftarrow \theta_i + \varsigma \nabla_{\theta_i} \mathcal{J}(\bm{\mu}_i)
\end{equation}
where $\varsigma \in (0,1]$ denotes the learning rate of the online actor network. 

\textbf{Critic}: The critic network is designed to approximate the Q-value function to assess the expected discounted cumulative reward by taking the global observations and joint actions as input. Each agent holds a separate critic network, estimating the Q-value function $Q_i\big(\mathbf{S},\mathbf{A}\big)$ parameterized by $\psi_i$.

To get a better approximation performance, the centralized action-value function $Q$ is updated as \cite{mnih2015human}:
\begin{equation}
    \label{critic_loss}
    \begin{aligned}
        \mathcal{L}(\psi_i) = \mathbb{E}_{\mathbf{S}, \mathbf{A}, \mathbf{R}, \mathbf{S}'} \Big[ \big(&Q_i(\mathbf{S}, \mathbf{A}) - y_i \big)^2 \Big],\\
        y_i = R_i + \beta \bar{Q}_i (\mathbf{S}', \mathbf{A}') &\Big|_{\mathbf{A}' = \big\{\bm{\mu}'_j (\bm{o}_j)\,|\,j\in\mathcal{K}\,\text{or}\,\mathcal{K}_1\big\}},
    \end{aligned}
\end{equation}
where $\bar{Q}_i$ is the target critic network parameterized by $\psi'_i$, and $\bm{\mu}'_j$ is the target policy with delayed parameter $\theta'_j$. Similarly, the critic network is also updated based on a batch of sampled experiences from the replay buffer $\mathcal{D}$, and the parameter of its online critic network is updated by 
\begin{equation}
    \label{critic_para_update}
    \psi_i \leftarrow \psi_i - \varsigma \mathcal{L}(\psi_i).
\end{equation}
It is noted that the parameters of the target actor network and critic network of agent $i$ are then updated by making them slowly track the learned online networks, i.e.
\begin{equation}
    \label{target_actor_critic_para_update}
    \begin{aligned}
        \theta_i' &\leftarrow \tau \theta_i + (1 - \tau) \theta_i', \\
        \psi_i &\leftarrow \tau\psi_i - (1-\tau)\psi_i'
    \end{aligned}
\end{equation}
where $\tau$ is the update rate of the target networks \cite{guo2022multi}.

\textbf{Environment}: For the short-timescale problem, during a time slot, each agent first extracts its partial observation or state from the environment, and its actor generates the action to the environment. After the transmission process in Section \ref{Downlink_transmission}, an immediate partial reward $r$ is obtained and fed back to the actors. Then, the environment is updated, leading to updated observations or states for the actors. The long-timescale problem has a similar procedure, while the difference lies in the update of the environment. When the trajectory control action $\bm{a}^\text{T}$ is generated, the UAV movement will be executed during the next $N_l$ short-timescale time slots with the same velocity from the action $\bm{a}^\text{T}$.  
    
\textbf{Replay Buffer}: During the early training process, the transition of each time step $(\mathbf{S}, \mathbf{A},\mathbf{R}, \mathbf{S}')$ is stored in this replay buffer. After the number of transitions in the buffer has exceeded a predefined limit, the actor-critic network samples a batch of transitions as experiences to assist the training.

The pseudo code of our TTS-MADDPG algorithm is summarized in Algorithm \ref{alg:two_timescale_maddpg}. The framework comprises a main loop, which contains the training for both long-timescale and short-timescale tasks. To be specific, it begins with initialization (Lines 1–4), setting up actor-critic networks, target networks, replay buffers, and hyperparameters. Based on the $\epsilon$-greedy policy, each user scheduling agent selects the action at each short-timescale time slot (Lines 8-10), while each trajectory control agent generates the action at each long-timescale time slot (Lines 11-16). The environment processes joint actions, generates reward and next state, and the transitions are stored in replay buffers (Lines 17–19). Policy updates occur when buffer sizes exceed thresholds, with short-timescale updates (Lines 20–26) and long-timescale updates (Lines 27–33).

\section{Numerical Results}\label{section_numerical_results}
In this section, we analyze the performance of the proposed TTS-MADDPG algorithm through the simulated numerical results.

\subsection{Simulation Settings}
We consider a circular area with a radius of 500m, where the G-UEs are uniformly distributed. Unless otherwise specified, we set the $M = 60$ G-UEs, $K_1 = 4$ U-UAVs, and $K_0 = 1$ T-UAV. The safe zone center of T-UAV is the same as the center of the disaster area, with a radius of 100m. For each U-UAV, according to Fig. \ref{System_model}, its safe zone center is located at $(\pm 200\,\mathrm{m},\pm200\,\mathrm{m})$ with a radius of 50m. Table \ref{tab:simulation_setting} summarizes the default settings for the environment and algorithm \cite{al2014optimal,mozaffari2016unmanned}. Both short-timescale and long-timescale tasks shared the same values for learning rate of actor and critic, epsilon greedy policy parameter, discounting factor, and batch size. Regarding the neural network structure for the actor and critic, we utilize two layers of gated recurrent unit (GRU) and four fully connected (FC) layers. All experiments were conducted on a single node equipped with NVIDIA RTX-2080Ti GPU (32GB memory) and Intel Skylake CPU. The software environment includes Python 3.6.5, PyTorch 1.10.0, and CUDA 11.8. The training simulation for each method has 10 independent runs with different random seeds, with each run of 1000 episodes. The testing simulation for each method has 10 independent runs with different random seeds, with each run of 100 episodes. For the proposed TTS-MADDPG algorithm, one full training run of 1000 episodes takes approximately 3 hours and 40 minutes, with an average simulation speed of about 4.5 episodes per minute.
\begin{table}[!ht]
    \centering
    \renewcommand{\arraystretch}{1.2} 
    \caption{Table of Simulation Settings}
    \label{tab:simulation_setting}
    \begin{tabular}{|c|c|}  
        \hline
        \textbf{Parameters (Notation)} & \textbf{Value} \\
        \hline
        Constants for LoS probability ($a$, $b$) & 11.95, 0.136\\
        \hline
        Height ($H_t$, $H_u$) & 200 m, 100 m \\
        \hline
        Carrier Frequency ($f_t$, $f_u$) & 2.6 GHz, 700 MHz  \\
        \hline
        Transmission Power ($P_{k_0}$, $P_{k_1}$) & 24 dBm, 14 dBm \\
        \hline
        Number of Antennas ($A_t$, $A_u$) & 32, 16 \\
        \hline
        Bandwidth ($B_{k_0}$, $B_{k_1}$) & 100 MHz, 20 MHz \\
        \hline
        Short-Timescale Time Slot Length ($T$) & 30 ms \\
        \hline
        Packet Generation Poisson Parameter ($\lambda$) & 4 \\
        \hline
        Packet Drop Latency ($N_\text{con}$) & 10 \\
        \hline
        Size of Packet ($N_p$) & 0.3 Mbits \\
        \hline
        Association Number for T-UAV  & 20 \\
        \hline
        Association Number for U-UAV  & 10 \\
        \hline
        Scheduling Number for T-UAV ($\mathcal{C}^\text{t}_\text{scd}$)& 8 \\
        \hline
        Scheduling Number for U-UAV ($\mathcal{C}^\text{u}_\text{scd}$)& 4 \\
        \hline
        Dimensional Velocity for U-UAV ($v_d^\text{max}$) & 10 m/s \\
        \hline
        Moving Velocity of G-UEs ($v_w$) & 5 m/s \\
        \hline
        Episode Length ($L_e$) & 200 \\
        \hline
        Learning Rate of Actor & $10^{-4}$ \\
        \hline
        Learning Rate of Critic & $10^{-3}$ \\
        \hline
        Parameter of Epsilon Greedy Policy ($\epsilon$)&  0.4 \\
        \hline
        Discounting Factor ($\beta$)& 0.95 \\
        \hline
        Batch Size ($B_\text{S},B_\text{T}$) & 64, 64 \\
        \hline
    \end{tabular}
\end{table}

\subsection{Performance Analysis}
\begin{figure}[ht]
  \centering
  \includegraphics[width = 7.5cm]{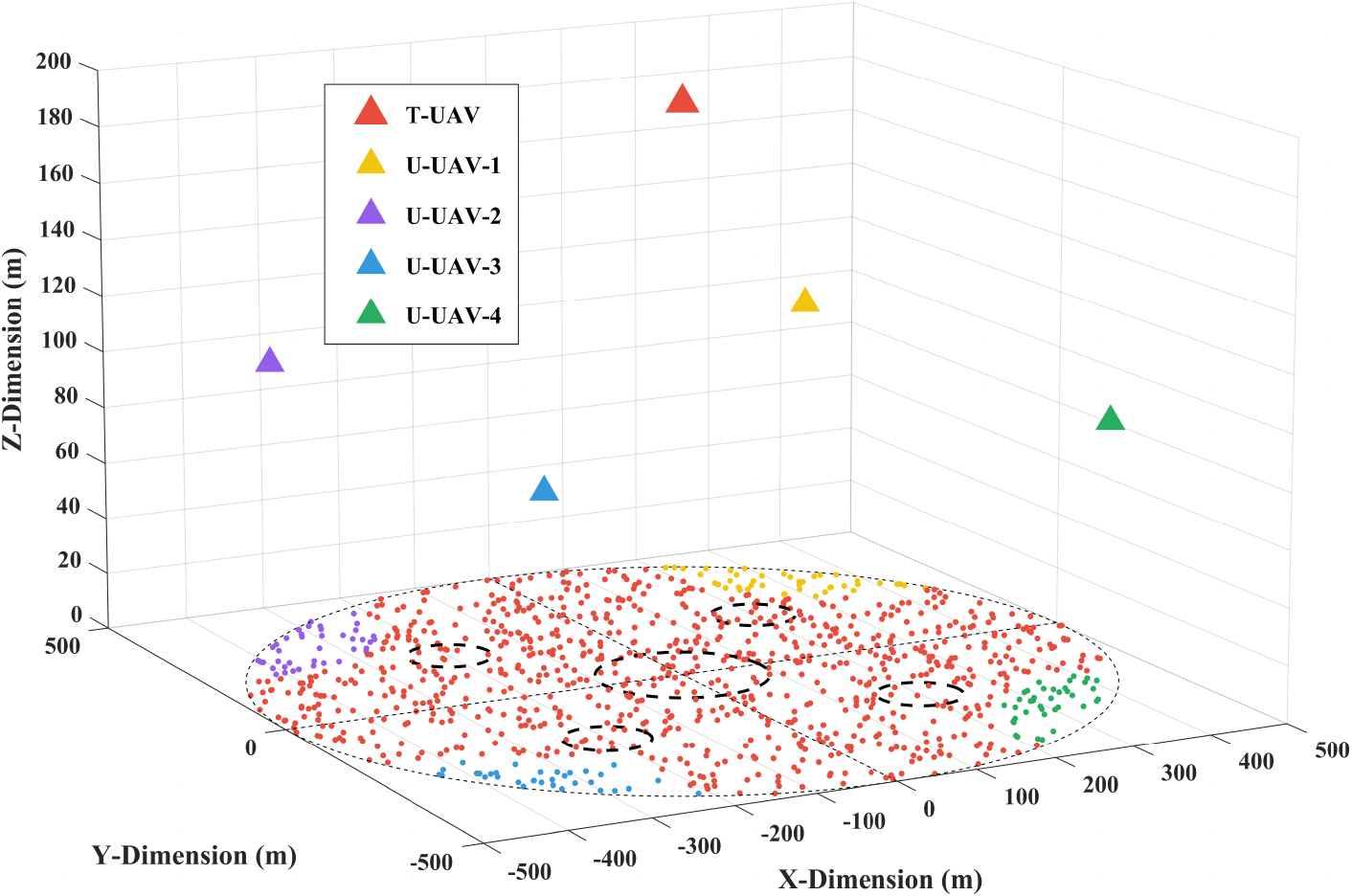}
  \caption{3D illustration of UAVs' coverage in post-disaster scenario.}
  \label{Fig_Coverage}
\end{figure}

Without loss of generality, we initialize the scenario with 1000 G-UEs and associate them with the corresponding UAVs. We randomly select $M = 60$ G-UEs (20 per T-UAV and 10 per U-UAV) from them for analysis in the following parts. The coverage of T-UAV and U-UAVs in such a post-disaster circular area is depicted in Fig. \ref{Fig_Coverage}.
In this figure, the G-UEs are color-coded to indicate their association with the corresponding UAVs. The T-UAV mainly serves the cell-center G-UEs, while the U-UAVs mainly serve the cell-edge G-UEs. The black dashed circles depict the safe zones for each group of G-UEs. 

\begin{figure}[ht]
  \centering
  \includegraphics[width = 7cm]{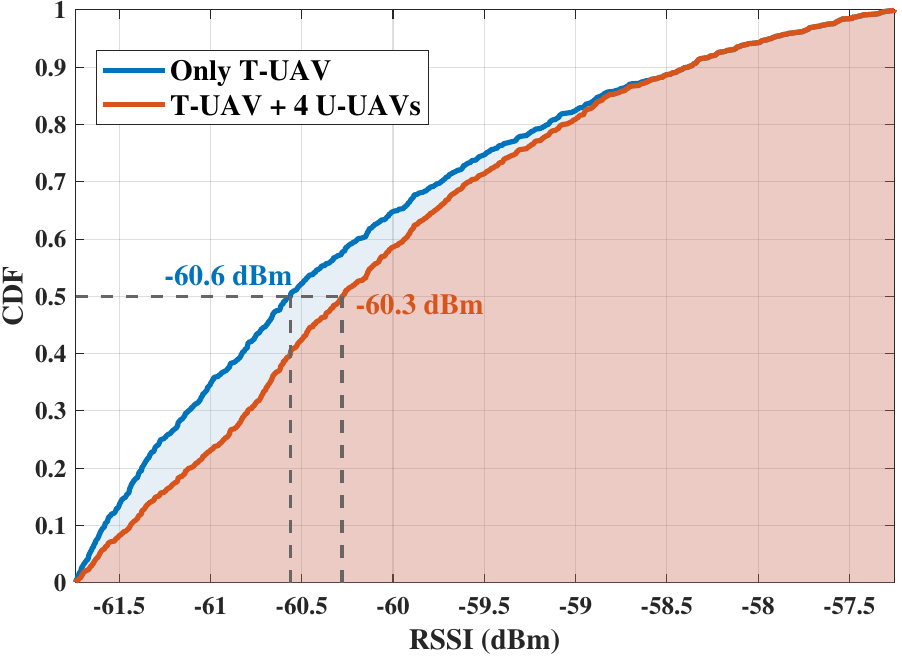}
  \caption{CDF of the RSSI for G-UEs under different UAV deployment strategies.}
  \label{Fig_CDF_RSSI}
\end{figure}

To demonstrate the benefits of employing U-UAVs for cell-edge G-UEs, we illustrate the comparison of RSSI cumulative distribution function (CDF) in Fig. \ref{Fig_CDF_RSSI}.
The blue curve represents the scenario with only the T-UAV providing service, while the red curve represents the results with the deployment of four additional U-UAVs. The results show the enhancement of RSSI by deploying multiple U-UAVs, with a shift in the distribution towards higher signal strengths. This improvement is attributed to the reduced communication distances and improved LoS conditions brought by the additional U-UAVs.

\begin{figure}[ht]
    \centering
    \subfigure[Training performance (Shade: 95\% confidence interval)]
    {\includegraphics[width=7.5cm]{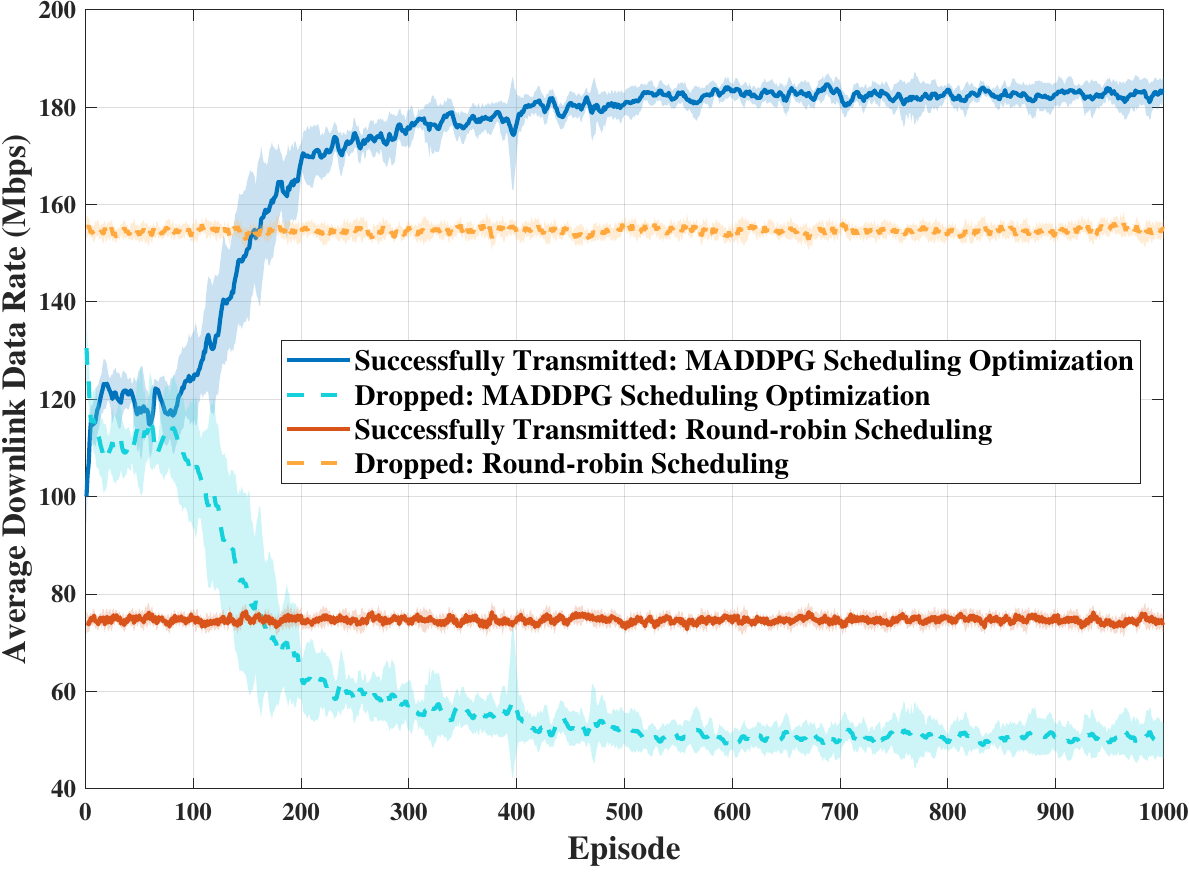}} \\
    \centering
    \subfigure[Testing performance for each UAV (Error bar: 95\% confidence interval)]    {\includegraphics[width=7.5cm]{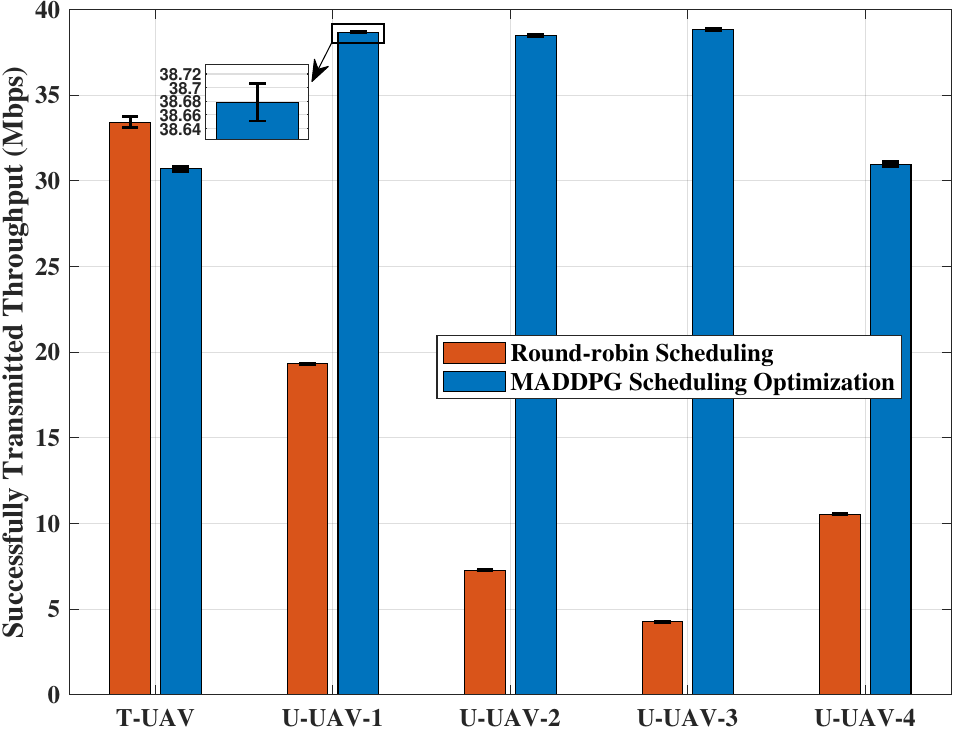}}
    \caption{Performance comparison of user scheduling between MADDPG and Round-robin methods under static G-UEs scenario.}
    \label{Fig_compare_scd_rr}
\end{figure}

We first evaluate the performance of user scheduling optimization under the static G-UEs scenario. We simulate the downlink throughput performance results of the proposed MADDPG-based user scheduling and Round-robin user scheduling during both the training and the testing phases, which is shown in Fig. \ref{Fig_compare_scd_rr}. 

Fig. \ref{Fig_compare_scd_rr}(a) shows the average downlink throughput during training. We can observe that the Round-robin user scheduling solution maintains a relatively constant throughput performance, with approximately 75 Mbps and 155 Mbps for the dropped and successfully transmitted throughput, respectively. In contrast, the MADDPG-based user scheduling solution increases the successfully transmitted throughput, converging at around 180 Mbps after about 300 episodes, which is about 140\% higher than the Round-robin user scheduling method. These trends highlight the agents' capability to obtain the optimal scheduling decisions based on MADDPG and show the limitations of the Round-robin method in such complicated scenarios.

\begin{figure*}[!ht]
  \centering
  \includegraphics[width=\textwidth]{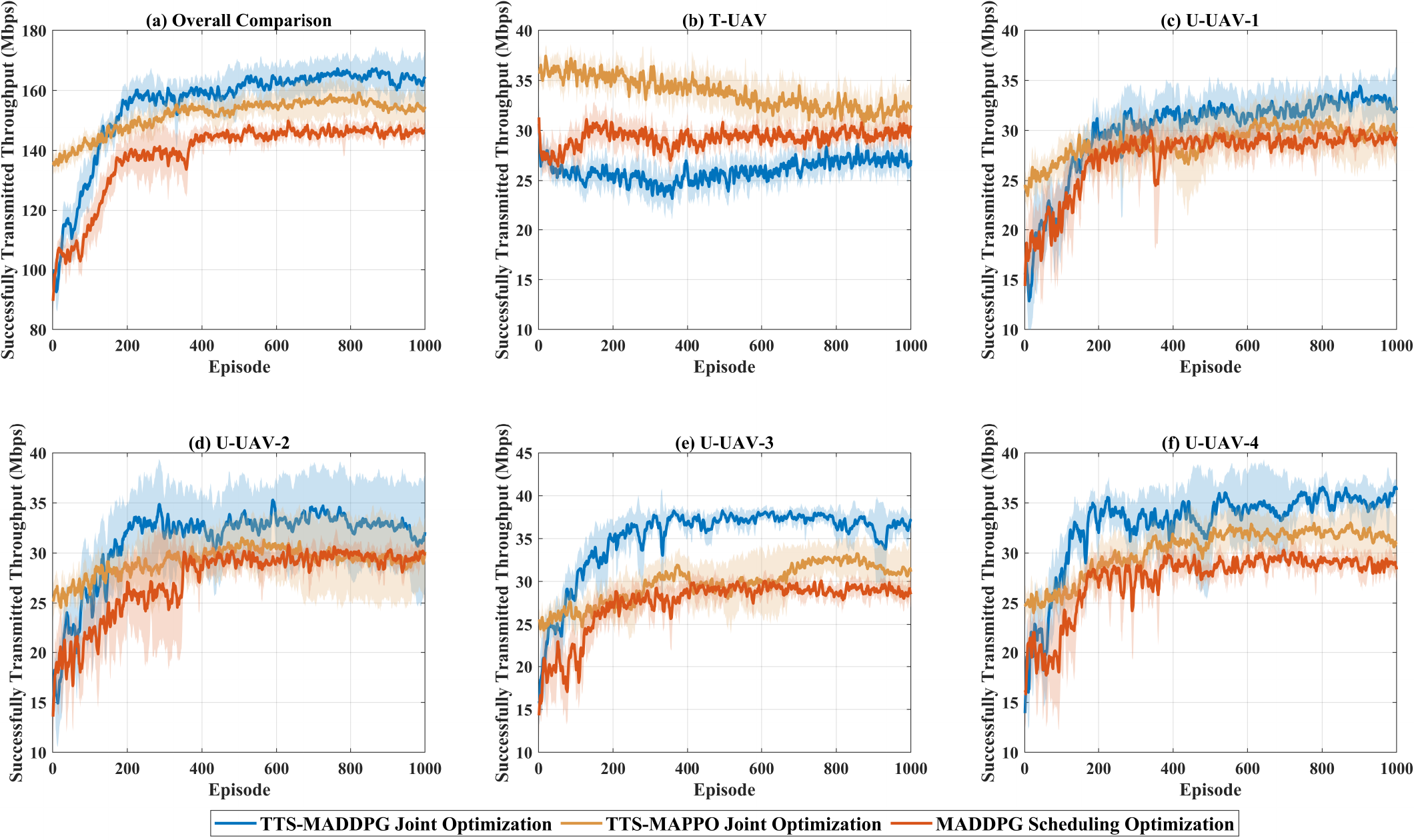}
  \caption{Training performance comparison among the proposed TTS-MADDPG joint optimization, TTS-MAPPO joint optimization, and MADDPG-based scheduling optimization under mobile G-UEs scenario (Shade: 95\% confidence interval).}
  \label{Fig_compare_jo_scd}
\end{figure*}

\begin{figure}[ht]
  \centering
  \includegraphics[width = 7.5cm]{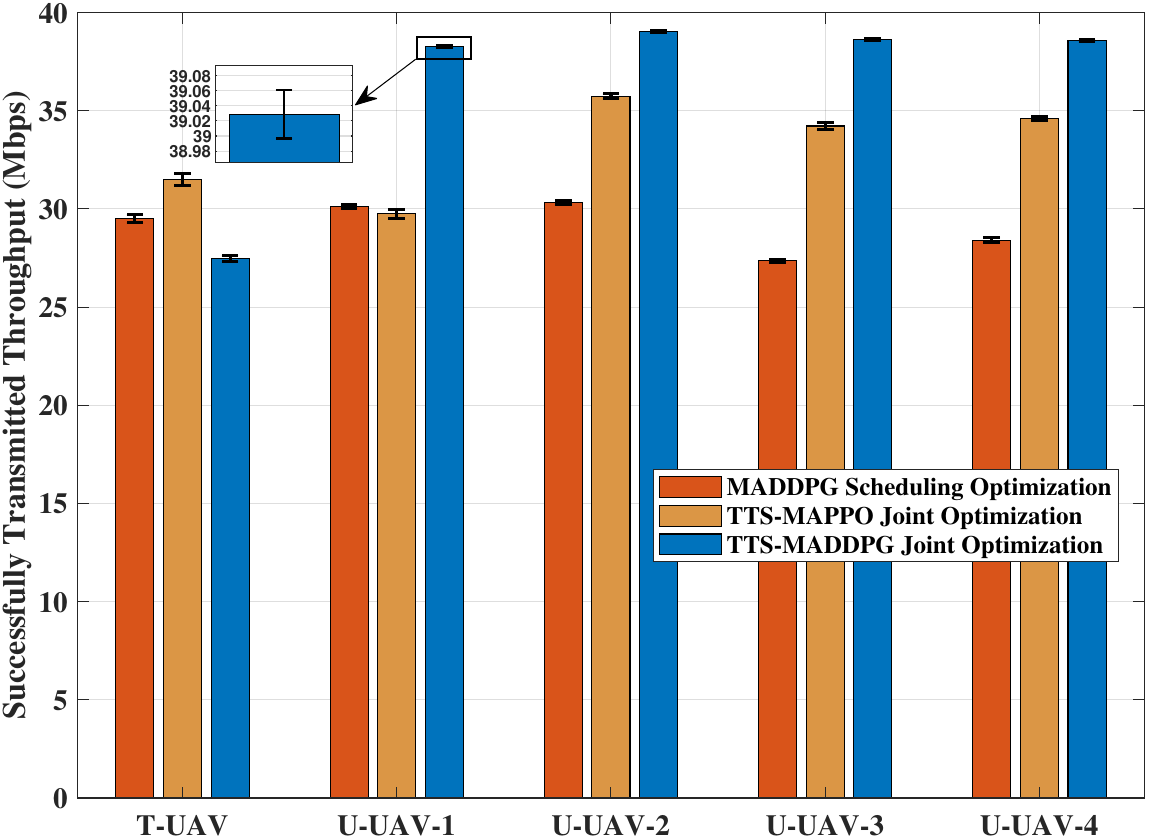}
  \caption{Testing performance comparison for each UAV among the proposed TTS-MADDPG joint optimization, TTS-MAPPO joint optimization, and MADDPG-based scheduling optimization under mobile G-UEs scenario (Error bar: 95\% confidence interval).}
  \label{Barfig_compare_jo_scd}
\end{figure}

During testing, to provide a thorough analysis, we illustrate the individual successfully transmitted throughput for each agent in Fig. \ref{Fig_compare_scd_rr}(b). It is worth noting that the MADDPG-based user scheduling method significantly outperforms the conventional Round-robin policy among all U-UAVs, but with a slight decrease compared to the Round-robin policy for T-UAV. 
For instance, the U-UAV-3 achieves about 38.68 Mbps compared to 19.31 Mbps with the Round-robin user scheduling solution, with an error bar of about 0.06 Mbps.
Therefore, the MADDPG-based user scheduling solution has been proven effective and robust in prioritizing IAB links due to asymmetric traffic demands.

We then proceed to evaluate the joint scheduling and trajectory optimization enabled by the proposed TTS-MADDPG algorithm. We have integrated PPO into our two-timescale framework, and introduced the TTS-MAPPO algorithm as one of the benchmarks \cite{schulman2017proximal,yu2022surprising}. Moreover, the MADDPG scheduling method is considered another benchmark as well to better show the performance gain brought by TTS-MADDPG algorithm.

Fig. \ref{Fig_compare_jo_scd} shows the training curves of the successfully transmitted throughput based on the proposed TTS-MADDPG joint optimization method, and its comparisons with two benchmarks. The overall throughput results are presented in Fig. \ref{Fig_compare_jo_scd}(a), where the TTS-MADDPG method outperforms the TTS-MAPPO method with higher throughput or faster convergence, with the averaged throughput converging up to 164.3 Mbps. The TTS-MADDPG-based joint optimization achieves a throughput gain of approximately 17.9 Mbps, representing a 12.2\% improvement over the 146.4 Mbps obtained by the MADDPG-based scheduling optimization.
Moreover, the throughput improvement brought by the TTS-MADDPG method is consistent across all U-UAVs in Figs. \ref{Fig_compare_jo_scd}(c)-(f), compared to other two benchmarks. For instance, in Fig. \ref{Fig_compare_jo_scd}(e), the converged curves of the TTS-MADDPG method reach around 38 Mbps, outperforming the 31 Mbps of TTS-MAPPO joint optimization and 28 Mbps of MADDPG-based scheduling optimization. It is worth noting that the throughputs from T-UAV exhibit completely different behaviors compared to the U-UAV in Fig. \ref{Fig_compare_jo_scd}(b), and the detailed interpretation towards this subfigure is provided below.
\begin{itemize}
    \item The TTS-MAPPO method obtains the highest throughput value for T-UAV, indicating its intention to lead the T-UAV to focus more on its scheduling decisions among the G-UEs rather than the U-UAVs, which will further negatively impact its overall performance on this IAB-enabled scenario.
    \item The TTS-MADDPG method exhibits a lower throughput value than the MADDPG scheduling method. This trend is mainly due to the more severe intra-cell interference to T-UAV's A2G transmission caused by the relatively closer distance between T-UAV and each U-UAV with optimized trajectories. Additionally, the optimized trajectories contribute to better channel capacities among U-UAV's A2G links, which requires T-UAV to sacrifice its A2G transmission and ensure sufficient data transmission on the A2A links to U-UAVs.
\end{itemize}

Similarly, with the well-trained models, we provide the individual successfully transmitted throughput for each UAV during the testing phase in Fig. \ref{Barfig_compare_jo_scd}. In this figure, the proposed TTS-MADDPG method still obtains the highest throughput value across all U-UAVs and a lower throughput on T-UAV, compared to two benchmarks. In addition, the error bars for the three methods are relatively narrow, indicating that the performance of each method is generally consistent and stable across runs. For example, with the TTS-MADDPG method, U-UAV-1 achieves a throughput of about 39.02 Mbps and obtains an error bar of less than 0.08 Mbps. These behaviors confirm that the differences between our proposed method and the benchmark results are statistically significant.

\begin{figure}[ht]
  \centering
  \subfigure[U-UAVs' optimized traces within one episode based on the proposed TTS-MADDPG algorithm]
  {\includegraphics[width=6.0cm]{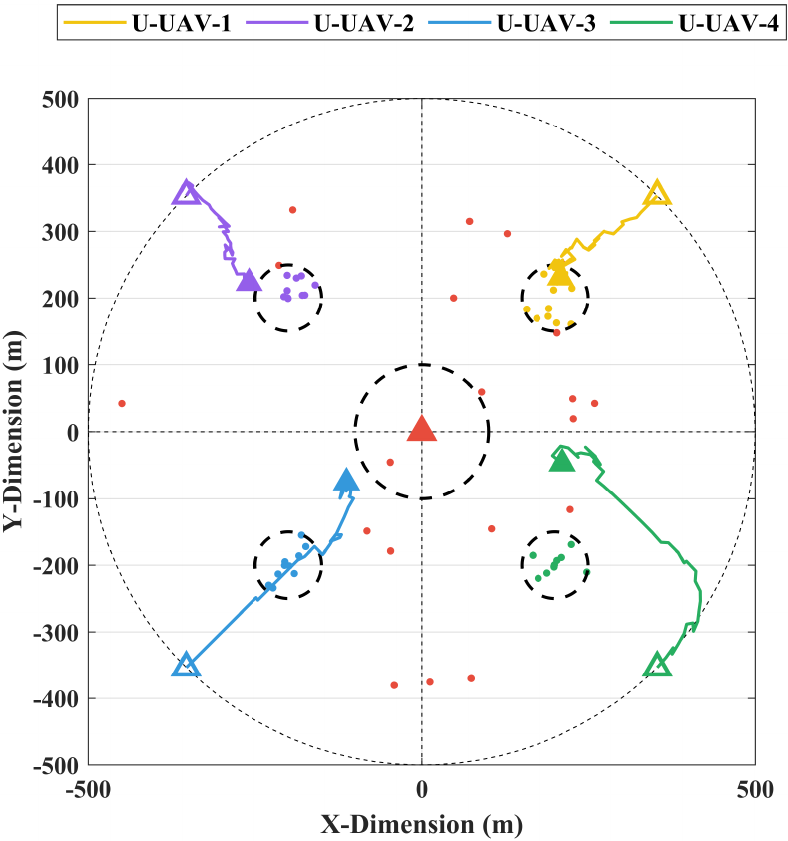}}
  \centering
  \subfigure[Time-slot-level downlink data rate within one episode (Shade: 95\% confidence interval).]
  {\includegraphics[width=7.2cm]{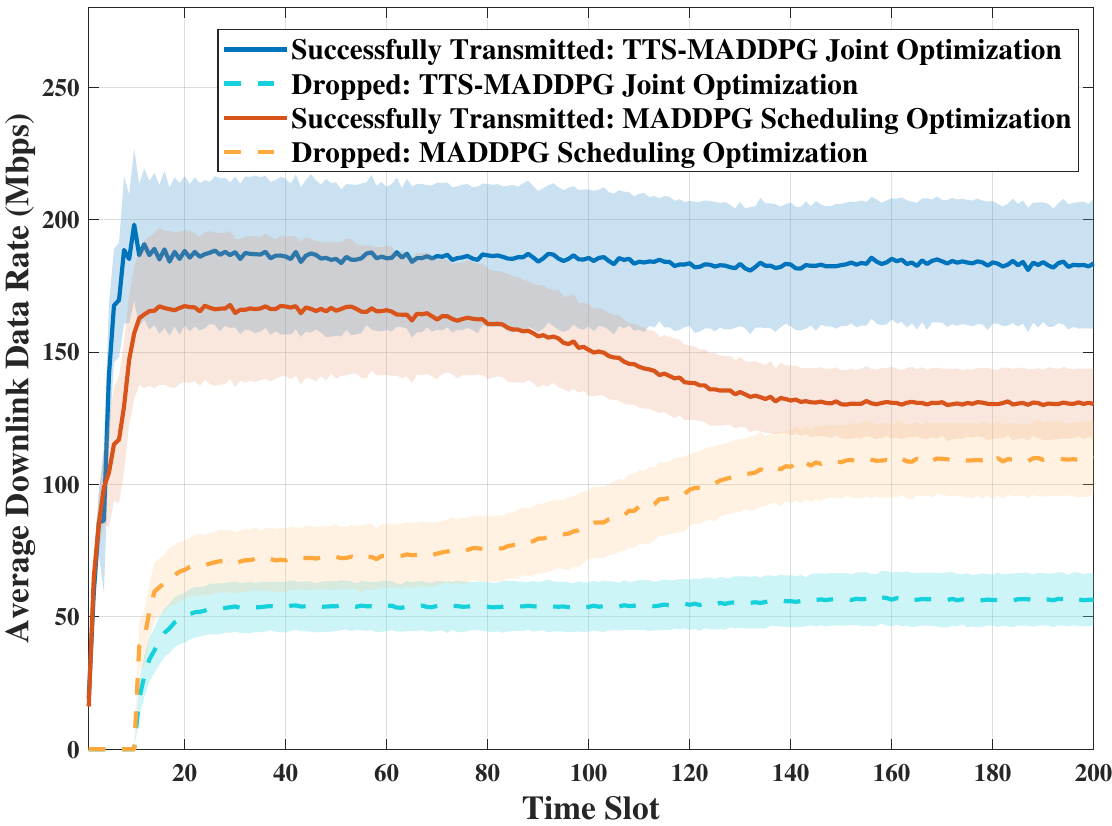}}
  \caption{Trace illustration and time-slot-level downlink data rate within one episode.}
  \label{fig:trace_throughput_slotlevel}
\end{figure}

To visually demonstrate the behaviors of each UAV, we illustrate the optimized U-UAVs' trace within an episode in Fig. \ref{fig:trace_throughput_slotlevel}(a). It is worth noting that UAVs' trace generally follows the overall movement direction of their associated groups of G-UEs. These behaviors confirm that the trained UAV agents have learned to adjust their trajectories to compensate for the performance degradation caused by G-UEs' motion. Moreover, we evaluate the time-slot-level behaviors of the overall dropped and successfully transmitted throughput within an episode during testing in Fig. \ref{fig:trace_throughput_slotlevel}(b). The MADDPG scheduling optimization method obtains improved and stable performance during the first 60 time slots, but then its performance deteriorates because of G-UEs' motion, with decreasing successfully transmitted data rate and increasing dropped data rate. In contrast, with the TTS-MADDPG joint optimization method, the successfully transmitted data rate remains stable at about 190 Mbps across nearly the entire 200 time slots, and the dropped data rate is maintained lower than 60 Mbps. These trends reflect the effectiveness of the proposed algorithm in optimizing U-UAV's trajectory to ensure seamless connectivity and stable throughput. Furthermore, the consistently stable successfully transmitted data rate across time slots validates that the proposed TTS-MADDPG algorithm effectively guarantees reliable communication quality, even in the presence of both intra-cell and inter-cell interference.

Therefore, the results in Fig. \ref{Fig_compare_jo_scd} and Fig. \ref{Barfig_compare_jo_scd} highlight the necessity of incorporating trajectory control into the optimization process and further validate the effectiveness of the proposed TTS-MADDPG algorithm over benchmarks such as the TTS-MAPPO algorithm.

\subsection{Ablation Study and Parameter Analysis}
In this subsection, we present the results of the ablation study and the parameter analysis on the proposed TTS-MADDPG algorithm.

We first conducted the ablation study on the employed neural network structure of the actor and critic, shown in Table \ref{tab:ablation}. 
\begin{table}[t]
    \centering
    \caption{Ablation study on the proposed TTS-MADDPG algorithm.}
    \label{tab:ablation}
    \vspace{0.1cm}
    \begin{tabular}{l|c|c}
    \hline
    \textbf{Method Variant} & \textbf{Throughput (Mbps)} & \textbf{Conv. Ep.}\\
    \hline
    \hline
    TTS-MADDPG (Full)          & \textbf{164.3}$\,\,\pm\,\,$\textbf{9.5}        & \textbf{204$\,$/$\,$1000}\\
    TTS-MADDPG w/o GRU         & 128.0$\,\,\pm\,\,$11.4                         & 785$\,$/$\,$2000 \\
    Benchmark (MADDPG Sched.)  & 146.4$\,\,\pm\,\,$3.4                          & 264$\,$/$\,$1000 \\
    \hline
    \end{tabular}
\end{table}
We make comparisons among the full model of TTS-MADDPG, the model without GRU layers of TTS-MADDPG, and the benchmark (MADDPG scheduling optimization), with the evaluation metrics of successfully transmitted throughput and convergence episode. The throughput value is calculated as the average converged value over the last 200 episodes, and the convergence episode is defined as the episode that first obtains or exceeds 95\% of the throughput value. The full neural network model of the proposed algorithm consists of GRU layers and FC layers. 
As the table shows, after deleting the GRU layers, the TTS-MADDPG algorithm experiences a throughput decrease of about 36.3 Mbps and requires more episodes to obtain clear convergence. These results further validate the importance and effectiveness of the GRU layers in accelerating convergence and increasing throughput.

\begin{figure}[ht]
  \centering
  \includegraphics[width = 7.5cm]{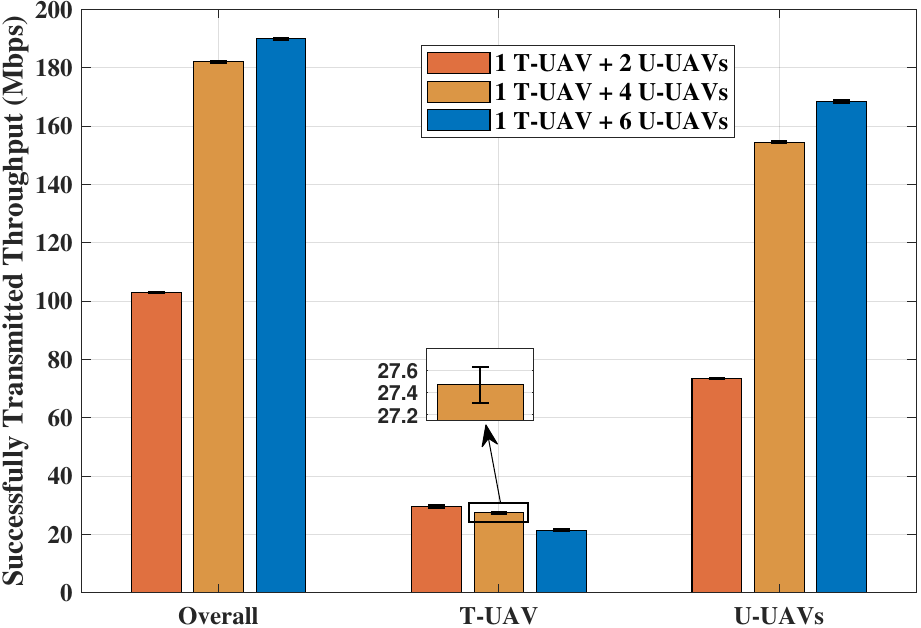}
  \caption{Testing performance for TTS-MADDPG joint optimizations given a different number of U-UAVs}
  \label{UAV_num_vary}
\end{figure}

We then conduct the parameter analysis on the number of U-UAVs deployed to support the communication service to edge G-UEs. We analyze the cases with 1 T-UAV and [2, 4, 6] U-UAVs, and evaluate their performance differences in the successfully transmitted throughput, which are shown in Fig. \ref{UAV_num_vary}.

In Fig. \ref{UAV_num_vary}, the overall throughput and the U-UAVs' throughput increase significantly as the number of U-UAVs grows, while the T-UAV throughput remains relatively low and slightly decreases. However, as the number of U-UAVs increases, the incremental throughput gain diminishes. Specifically, the increase from 2 to 4 U-UAVs contributes to a throughput gain of about 79 Mbps, while the increase from 4 to 6 U-UAVs only yields an additional 8 Mbps. Meanwhile, the error bars for these three settings are relatively small, which indicates the good generalization capability of the proposed algorithm towards different numbers of agents and various environments.

\section{Conclusion}\label{section_conclusion}
In this paper, we proposed an IAB-enabled heterogeneous UAV-based network for emergency communications, where U-UAVs are utilized to enhance the performance of cell-edge G-UEs during post-disaster activities. Then, we formulated a joint user scheduling and trajectory control optimization problem considering the asymmetric traffic demands in IAB and G-UEs' mobility, aiming to maximize the downlink successfully transmitted throughput. Finally, we developed a TTS-MADDPG algorithm based on the CTDE framework to solve the problem in a distributed manner, where user scheduling is optimized at the short-timescale time slot for both T-UAV and U-UAVs, and trajectory control is performed at the long-timescale time slot for each U-UAV. Extensive simulations validate the optimization effectiveness of the proposed TTS-MADDPG algorithm, which outperforms the TTS-MAPPO algorithm and MADDPG scheduling method in terms of successfully transmitted throughput. The ablation study and parameter analysis further confirm the effectiveness and good generalization capability of the proposed algorithm. In future work, we plan to integrate other NTN platforms, such as HAPs or satellites, into the emergency network and explore the feasibility and performance of the hierarchical MADRL algorithm.

% ————————————————————————————————————————————————%
% if have a single appendix:
%\appendix[Proof of the Zonklar Equations]
% or
%\appendix  % for no appendix heading
% do not use \section anymore after \appendix, only \section*
% is possibly needed

% use appendices with more than one appendix
% then use \section to start each appendix
% you must declare a \section before using any
% \subsection or using \label (\appendices by itself
% starts a section numbered zero.)
%

% \appendices
% \section{Proof of the First Zonklar Equation}
% Appendix one text goes here.

% % you can choose not to have a title for an appendix
% % if you want by leaving the argument blank
% \section{}
% Appendix two text goes here.

% use section* for acknowledgment
% \section*{Acknowledgment}

% The authors would like to thank...

% Can use something like this to put references on a page
% by themselves when using endfloat and the captionsoff option.
\ifCLASSOPTIONcaptionsoff
  \newpage
\fi

% trigger a \newpage just before the given reference
% number - used to balance the columns on the last page
% adjust value as needed - may need to be readjusted if
% the document is modified later
%\IEEEtriggeratref{8}
% The "triggered" command can be changed if desired:
%\IEEEtriggercmd{\enlargethispage{-5in}}

% references section

% can use a bibliography generated by BibTeX as a .bbl file
% BibTeX documentation can be easily obtained at:
% http://mirror.ctan.org/biblio/bibtex/contrib/doc/
% The IEEEtran BibTeX style support page is at:
% http://www.michaelshell.org/tex/ieeetran/bibtex/
\bibliographystyle{IEEEtran}
% argument is your BibTeX string definitions and bibliography database(s)
\bibliography{IEEEabrv,ref}
\end{document}